\documentclass[12pt,letterpaper]{article}

\usepackage[T1]{fontenc}
\usepackage[utf8]{inputenc}
\usepackage{amsmath,amssymb,amsthm,mathtools}
\usepackage{mathpazo}
\usepackage{geometry}
\usepackage{setspace,microtype,enumitem}
\usepackage{sectsty}
\usepackage[round,authoryear]{natbib}
\usepackage{etoolbox,needspace}
\usepackage{hyperref}
\usepackage{xcolor}
\usepackage{tikz}
\usetikzlibrary{arrows.meta,calc}
\usepackage{caption}
\usepackage{placeins}
\definecolor{benchmarkblue}{HTML}{205B7D}
\definecolor{benchmarkorange}{HTML}{C95B39}
\definecolor{benchmarkpurple}{HTML}{8055A6}
\definecolor{benchmarkteal}{HTML}{258589}
\definecolor{benchmarkgray}{HTML}{96A1A8}
\tikzset{
  benchmark axis/.style={draw=benchmarkgray,line width=.55pt,-{Stealth[length=4pt]}},
  benchmark guide/.style={draw=benchmarkgray,line width=.4pt},
  benchmark high/.style={draw=benchmarkblue,line width=1.15pt,line cap=round,line join=round},
  benchmark low/.style={draw=benchmarkorange,line width=1.05pt,dash pattern=on 4pt off 2.5pt,line cap=round,line join=round},
  benchmark critical/.style={draw=benchmarkpurple,line width=1.15pt,dash pattern=on 0.9pt off 2.6pt,line cap=round,line join=round},
  benchmark text/.style={font=\footnotesize,inner sep=1.5pt},
  benchmark title/.style={font=\small\bfseries,align=center,inner sep=1pt}
}
\pgfmathdeclarefunction{benchmarkP}{2}{%
  \pgfmathparse{0.5*#2*#1*#1-asin(sqrt(#1))*pi/180+sqrt(#1*(1-#1))}%
}

\sectionfont{\fontsize{15}{15}\selectfont}
\subsectionfont{\fontsize{14}{14}\selectfont}
\setlist[enumerate]{label=(\roman*),leftmargin=*,itemsep=3pt,topsep=4pt}
\allowdisplaybreaks[2]

\newtheorem{assumption}{Assumption}
\newtheorem{theorem}{Theorem}
\newtheorem{proposition}{Proposition}
\newtheorem{lemma}{Lemma}
\newtheorem{corollary}{Corollary}
\theoremstyle{definition}
\newtheorem{definition}{Definition}
\theoremstyle{remark}

\BeforeBeginEnvironment{theorem}{\Needspace{6\baselineskip}}
\BeforeBeginEnvironment{proposition}{\Needspace{6\baselineskip}}
\BeforeBeginEnvironment{corollary}{\Needspace{6\baselineskip}}
\BeforeBeginEnvironment{lemma}{\Needspace{6\baselineskip}}
\BeforeBeginEnvironment{definition}{\Needspace{6\baselineskip}}

\newcommand{\R}{\mathbb R}
\newcommand{\E}{\mathbb E}
\newcommand{\Ccal}{\mathcal C}
\newcommand{\Pcal}{\mathcal P}
\newcommand{\Mcal}{\mathcal M}
\newcommand{\Xcal}{\mathcal X}
\newcommand{\dd}{\,\mathrm d}
\DeclareMathOperator*{\argmax}{arg\,max}
\DeclareMathOperator*{\argmin}{arg\,min}
\DeclareMathOperator{\dist}{dist}

\hypersetup{
  colorlinks,
  linkcolor=cyan,
  citecolor=cyan,
  urlcolor=magenta,
  pdftitle={Building a Sovereign Benchmark: The Limits of Private Coordination},
  pdfauthor={Feng Dong and Fan Wu}
}

\title{\textbf{Building a Sovereign Benchmark}\\[0.3em]
\textbf{The Limits of Private Coordination}\thanks{%
Feng Dong acknowledges financial support from the National Natural Science Foundation of China (72525004). Fan Wu acknowledges financial support from the National Natural Science Foundation of China (72603022). The usual disclaimer applies.}}

\author{
Feng Dong\thanks{dongfeng@sem.tsinghua.edu.cn;
School of Economics and Management, Tsinghua University.}
\and
Fan Wu\thanks{fanwu@phbs.pku.edu.cn;
HSBC Business School, Peking University.}
}
\date{October, 2026}

\begin{document}
\maketitle

\begin{abstract}
Public debt supplies liquidity and collateral services only if institutions coordinate on using the same claim. We study this coordination when use carries a recurring cost and a benefit that rises with other users. Above a quality threshold, two positive-adoption equilibria exist, and under best responses the initial user base decides between collapse and high adoption. Under logit choice, as the population grows and noise vanishes, the long-run distribution selects high adoption only above a strictly higher threshold. 
Debt issuance improves
market depth but eventually dilutes backing per claim. This
trade-off creates a range of debt levels supporting the liquidity premium of public debt in macroeconomic models and a narrower range supporting its long-run selection. At the boundary of the wider range, the premium is discontinuous. A matching transfer that corrects the
adoption externality makes stationary selection maximize network surplus.
Transfers can also help launch a network without permanent support.


\end{abstract}
\baselineskip18.5pt\newpage
\section{Introduction}
\label{sec:introduction}

Government bonds provide services beyond their cash payoff. Banks and
dealers use them as collateral to obtain short-term funding through
repurchase agreements, or repos. Institutions can also post them to meet
initial margin requirements where they are eligible, take positions in
them to hedge interest-rate risk, or borrow them to fulfill securities
delivery obligations. These activities depend on counterparties that
are willing and able to transact in the same claim. A dealer's ability
to finance a position, for example, is more valuable when a broader set
of lenders accepts the security as collateral.  Common use therefore
creates opportunities that the claim's promised repayment alone does
not provide. The service is produced by the users, not owned by the asset.

These services are central to the macroeconomic role of public debt.
By supplying liquidity, public debt can relax private financing
constraints \citep{Woodford1990,HolmstromTirole1998,AiyagariMcGrattan1998}.
The resulting liquidity premium shapes the optimal quantity of
government debt \citep{AngeletosCollardDellas2023,BassettoCui2024}.
That premium is one reason the government's borrowing rate can sit below the growth rate, which changes the fiscal cost of carrying debt \citep{Blanchard2019,Reis2022,Kocherlakota2023}.
 Liquidity and safety
services also matter for the valuation of the outstanding debt stock
\citep{KrishnamurthyVissingJorgensen2012,
JiangLustigVanNieuwerburghXiaolan2024,BrunnermeierMerkelSannikov2024}. In these models, the liquidity of public debt is taken as given: the claim is fully liquid, or a demand for it is specified directly.
Whether these services persist as debt expands therefore matters for
both private finance and public borrowing. This question is particularly
relevant for U.S. Treasuries given the projected growth in public
debt.\footnote{In its February 2026 baseline, the Congressional Budget
Office projects federal debt held by the public to rise from 101 percent
of GDP in 2026 to 120 percent in 2036 \citep{CBO2026}.}

In practice, sustaining common use requires institutions to maintain valuation systems, collateral-management arrangements, and settlement procedures. In this paper, adopting a government security means
maintaining the capacity to use it in these activities, rather than
simply buying and holding it. Each institution weighs the ongoing cost
of that capacity against the opportunities supplied by other users.
An established reference consequently benefits from its installed base,
while reliable repayment alone may not justify participation in a
sparsely used claim. A related coordination force arises for reference
rates: \cite{DuffieStein2015} explain how the liquidity of LIBOR-linked
markets encouraged participants to continue trading instruments tied to
LIBOR. We examine this participation margin for government securities,
treating their common use as an equilibrium outcome. Our focus is on when the network supplying these services can sustain itself, and which adoption outcome prevails when institutions repeatedly reconsider their participation. 

To study this coordination problem, we consider institutions choosing
between using a government claim as their reference and retaining an
outside convention. Institutions differ in their recurring
implementation costs. The benefit of adoption rises with the share
of other users and with benchmark quality, which combines market depth
with the fraction of the real payoff retained by a holder. We first
hold quality fixed to study adoption, and then examine how debt supply
affects quality. An equilibrium requires the share of institutions
willing to adopt to equal the share actually using the reference. The model gives a microfoundation for when the service exists, not for how much each unit of it is worth.

The first result characterizes these equilibria. If nobody uses the
reference, no institution wants to pay its implementation cost alone, so
zero adoption is always an equilibrium. When quality is too low, it is
the only equilibrium. Once quality reaches a threshold, a positive
network can sustain itself. Above that threshold, there are exactly two
positive equilibria, one with fewer users and one with more users, in
addition to zero adoption. The two positive equilibria meet at the
threshold and disappear below it. Thus sufficiently good quality makes
a functioning network possible, but does not make its adoption
inevitable.

We next specify how institutions revise their decisions. Each institution
receives opportunities to reconsider adoption independently of the others.
Between opportunities, its action remains unchanged. When an opportunity
arrives, it compares the benefits of using the current network with its
implementation cost. We use a logit choice rule: the more profitable
action is more likely, but either action can be chosen when revision
noise is positive. As this noise vanishes, choices approach best response.

Without revision noise, the adoption network follows a
deterministic adjustment path. The zero-adoption and high-adoption equilibria are locally stable, whereas the low-adoption equilibrium is unstable.
An initial user base below the low-adoption equilibrium shrinks;
one above it moves toward the high-adoption equilibrium. The low-adoption equilibrium is therefore the
critical mass separating eventual abandonment from convergence to the
high-adoption equilibrium. An established
reference can consequently persist under conditions in which a smaller
initial user base would fail to grow.

With a finite population and positive revision noise, the long-run
frequencies of adoption are described by a unique stationary distribution. This distribution is a Gibbs measure: an adoption configuration
receives a weight that increases exponentially with its potential. The
potential is a summary of private incentives. When one institution
changes its action, the change in potential equals that institution's
payoff gain.  As the population grows and revision noise vanishes, the stationary
distribution concentrates on the equilibria with the highest potential. Only zero adoption and the high-adoption equilibrium are candidates for the potential maximizer.

We then characterize when the high-adoption equilibrium wins this
potential comparison. The high-adoption equilibrium is the unique potential maximizer if and only if quality exceeds a threshold. This threshold lies strictly above the quality needed for a positive equilibrium to exist, and strictly below twice that level. The reason is
that sustaining a network at a given size and overcoming the private
incentives against adoption at smaller sizes are different requirements.
At the existence threshold, the positive equilibrium still has lower
potential than nonadoption. Further improvements in quality eventually
reverse that ranking. Between the two thresholds, an established benchmark can persist, but it is not where a noisy economy spends most of its time.

To translate these results into debt supply, we suppose that benchmark
quality first rises and then falls with the amount of eligible debt.
A small issue offers too little market depth. A larger issue improves
depth, but eventually spreads the resources backing the claim across
more promised payments. The debt levels supporting a positive equilibrium therefore form
a feasibility corridor. The stronger requirement for stationary
selection produces a narrower coordination corridor. A debt level may support an established network
without placing that network in the long-run selected regime. 
\cite{BrunnermeierMerkelSannikov2024} and \cite{HeKrishnamurthyMilbradt2019} also find that debt size works in both directions, the first through a service flow spread over a larger stock and the second through market depth and rollover risk.

We next discuss welfare implications. The quality needed for positive net surplus is below the quality needed for a positive network to exist. Between those two, a network would create surplus, yet none can exist without support.
A policy exercise holds debt and backing fixed and examines a transfer
that rewards adoption in proportion to the existing network. An
institution considers the benefit it receives from joining, but not the
benefit its participation creates for other users. Under nondistortionary
transfer finance, a payment equal to this external benefit aligns the
potential with aggregate network surplus. The same stationary selection
mechanism then implements a surplus-maximizing adoption share. 

The transfer can also help launch a network without being needed
permanently. When benchmark quality is above the existence threshold,
a high-adoption equilibrium can sustain itself, yet a small initial
user base may fail to reach it. Under noiseless renewal, a matching
transfer that makes this initial network grow eventually raises
adoption above the critical mass required without support. The
government can then withdraw the transfer, and adoption converges
to the unsubsidized high-adoption equilibrium. This distinguishes
launching a network from maintaining it: temporary support helps
build the user base, while the network's own benefits sustain
participation once that base is sufficiently large.\footnote{\cite{CrouzetGuptaMezzanotti2023} find that a temporary cash shortage in India left electronic-payment adoption permanently higher.}
At fixed
benchmark quality, the intervention changes which equilibrium is
reached without requiring a permanent subsidy to maintain it.

Finally, a pricing extension separates the claim's cash payoff from the service supplied by its adoption network, the cash-flow plus service-flow split of \cite{BrunnermeierMerkelSannikov2024}. We follow an operating network along the high-adoption branch to the upper debt level at which positive equilibria cease to exist. Beyond that boundary only nonadoption remains, and the price falls by the discounted value of the lost service. \cite{ChenEtAl2025} document historical cases in which the dominant supplier of safe debt later lost that position. In the model that position can be lost without a jump in backing.

\paragraph{Related literature.}
A macroeconomics literature treats government debt as an asset that eases private liquidity constraints. \cite{Woodford1990}, \cite{HolmstromTirole1998}, and \cite{AiyagariMcGrattan1998} give public debt a role as private liquidity, or as a store of value when private markets are incomplete. \cite{AngeletosCollardDellas2023} and \cite{BassettoCui2024} study how much of that debt a government should supply when it carries a liquidity premium. \cite{BassettoCui2018} study fiscal backing when interest rates are low. \cite{Blanchard2019}, \cite{Reis2022}, and \cite{Kocherlakota2023} examine the fiscal arithmetic of rolling debt over when the interest rate is low relative to growth. \cite{DongSargentWangWang2026} show that persistent primary deficits can produce multiple steady states and select a low-interest, low-productivity equilibrium, so an interest rate below growth need not be a free lunch. \cite{KiyotakiMoore2019} let assets differ in how much of them can be resold in a period. \cite{JiangSargentWangYang2024} determine government debt capacity when debt has a convenience yield and the issuer can default. \cite{JiangSargentWang2026} show that financing accumulated deficits with short-term debt lets later governments carry out an optimal tax plan even when inherited debt is high. 

Other work studies the same debt as a safe asset. \cite{FarhiMaggiori2018} show that issuing more reserve assets exposes the issuer to a confidence crisis. \cite{CaballeroFarhi2018} study a shortage of safe assets that produces a deflationary trap at the zero lower bound, where more public safe debt raises output. \cite{JiangKrishnamurthyLustig2024} connect the safety of dollar debt to global financial conditions. \cite{BrunnermeierMerkelSannikov2024} price a safe asset as the sum of its cash flows and a service flow from retrading, and the value of that service can fall as more debt is issued. \cite{ChenEtAl2025} find that dominant suppliers of safe debt have lost that role when fiscal backing weakened. These papers take the liquidity of public debt as given, or they derive it from the asset's payoff. We derive it from institutions' decisions to use the claim.

A related line of work makes the acceptability of an asset the result of individual choices. \cite{KiyotakiWright1989} have an asset become a medium of exchange because others accept it. \cite{LesterPostlewaiteWright2012} let agents pay to recognize an asset: recognition by more agents makes it more liquid, which raises its price and the return to recognition, and multiple equilibria can result. In \cite{AsriyanFuchsGreen2019}, adverse selection and resale make liquidity today depend on expected liquidity tomorrow, and sunspots select among self-fulfilling equilibria. \cite{CuiRadde2020} obtain the liquidity of private claims from investor participation in a search market, with government bonds filling the gap when that participation is low. \cite{AtkesonEisfeldtWeill2015} study banks' entry into and exit from an over-the-counter market; after large negative shocks, dealers can exit more than is socially optimal.

International macroeconomics asks the same question about currencies and safe assets. \cite{MatsuyamaKiyotakiMatsui1993} have national currencies circulate only locally or as an international currency, and use an evolutionary approach to select among those equilibria. \cite{ChahrourValchev2022} show that saving in an asset and using it to finance trade reinforce each other, which produces several stable steady states, each with a different dominant asset. \cite{CoppolaKrishnamurthyXu2025} let liquidity provision and the supply of safe bonds drive currency use, with transitions between regimes that depend on safe-asset supply. For a single sovereign claim, the single-peaked cost distribution delivers a sharp threshold for a positive network to exist and a higher threshold for long-run selection.

The adoption model belongs to the literature on network externalities and
interdependent discrete choices. \cite{KatzShapiro1985} and
\cite{FarrellSaloner1985} establish how compatibility and the choices of
other users shape adoption incentives.
\cite{Rohlfs1974} has the same critical-mass logic: zero use is stable, and the smaller positive equilibrium is not. \cite{FarrellSaloner1986} show that the users already in place can lock in the existing standard.
\cite{BrockDurlauf2001} study
self-consistent aggregate choices and multiplicity with social
interactions.\footnote{An additive, fixed-heterogeneity formulation is
\cite{GordonEtAl2009}, who analyze binary adoption with a distribution of
individual willingness to pay and a benefit that increases with the
adoption share.}  
We impose a thin lower tail and a single-peaked cost density to
obtain a sharp existence threshold, and compare that threshold with the
one generated by stationary logit selection.

Our adjustment mechanism uses tools from potential games and
evolutionary dynamics. The exact-potential property follows the framework
of \cite{MondererShapley1996}; \cite{Sandholm2001} and
\cite{Sandholm2009} develop potential games for large populations.
\cite{Blume1993} connects noisy individual revisions, invariant
distributions, and equilibrium selection. \cite{Sandholm2010} studies
large-population and small-noise limits of stationary distributions and
shows that selection can depend on the noisy response rule. The Gibbs
formula and the general potential-selection principle are therefore
methodological foundations, rather than new selection criteria. We apply
them to a population with fixed heterogeneous costs, prove the
approximation from finite action profiles to the continuum potential,
and use one revision protocol to distinguish installed-base dependence
on finite horizons from its stationary selection. The 
existence boundary does not depend on choosing logit as the revision rule;
the coordination boundary is specific to the maintained selection
mechanism.
\cite{KandoriMailathRob1993} and \cite{Young1993} select by rare mutations, and the expected wait grows exponentially with population size, so a finite horizon need not reach the stationary outcome. \cite{Ellison1993} shows that under uniform matching these waits are long enough for history to determine play, while interaction among a few neighbors shortens them. Adoption benefits here depend on aggregate use, which is the uniform case. \cite{CarlssonVanDamme1993} and \cite{MorrisShin2003} select from incomplete information about payoffs.
\cite{FrankelPauzner2000} study sectoral choice with external increasing returns, where workers switch only at random times. A Brownian shock that can make either sector dominant selects a unique path. Macro-finance models more often resolve multiplicity with a sunspot \citep{AzariadisKaasWen2016} or with a global game \citep{HeKrishnamurthyMilbradt2019}. The revision protocol used here keeps those equilibria and adds a criterion for where the economy spends its time, which is what separates the two thresholds.

A close sovereign-debt comparison is
\cite{HeKrishnamurthyMilbradt2019}. They determine safe-asset status
through investors' coordination across sovereign debt markets, with debt
size affecting both market depth and exposure to rollover risk. The
trade-off between depth and backing is therefore not specific to the
present paper. Here there is one eligible reference claim, its payoff
retention is a primitive, and institutions coordinate on using it rather
than on refinancing its issuer.
In \cite{ColeKehoe2000}, lenders coordinate on refinancing and a failed auction leads to default. \cite{AguiarEtAl2022} add issuance at depressed prices and sudden stops. In \cite{LorenzoniWerning2019}, fear of default raises interest rates, debt then grows faster, and default eventually follows.
Here adoption does not determine the claim's cash repayment, so the price drop at the edge of the feasibility corridor is a loss of service rather than a default. Our question is whether the use of that claim as a common
reference can be sustained, and how the set of supporting debt levels
changes when a stationary selection requirement is added to equilibrium
existence. The single-peaked quality assumption makes the trade-off between depth and retention explicit without modeling a rollover game.
Other financial-reference models study different benefits from common
use. \cite{DoepkeSchneider2017} explain a dominant unit of account through
noncontingent contracts and the reduction of relative-price risk along
credit chains. \cite{DuffieDworczakZhu2017} study informational benchmarks
that improve price transparency, search, and matching in over-the-counter
markets.
\cite{Yuan2005} models a benchmark security as a way to trade systematic risk apart from security-specific risk. \cite{DittmarYuan2008} find that sovereign bonds play this role for corporate bonds in emerging markets.
Our reference is a security around which institutions organize
their activities, not a published price signal; neither information
aggregation nor denomination choice drives the adoption equation.
\cite{KrishnamurthyVissingJorgensen2012} document the value of liquidity
and safety services in Treasury yields and their response to supply.
Our pricing extension takes the valuation of such a service as given.
Its content is the loss of that service when the positive-adoption
branch ends, rather than a new pricing formula or an empirical estimate
of convenience yields.

Finally, the corrective-transfer result is an application of the
stochastic evolutionary implementation approach of \cite{Sandholm2007}.
That paper shows how Pigouvian prices can produce efficient long-run
behavior under logit revisions despite limited information and multiple
equilibria. In our setting the correction takes the simple form of a
payment proportional to aggregate adoption. It aligns the adoption
game's potential with network surplus, without eliciting individual cost
reports.
\cite{DybvigSpatt1983} treat adoption externalities as a public good and note a role for intervention when one equilibrium Pareto-dominates the others.
\cite{BenignoFornaro2018} show that investment subsidies can lift an economy out of a stagnation trap by shifting growth expectations. A temporary matching transfer works here through the installed base: it raises adoption above the critical mass, after which the network sustains itself.
\cite{SakovicsSteiner2012} instead study how to target subsidies
across heterogeneous agents to achieve successful coordination at minimum
cost under strategic uncertainty. We do not solve that targeting problem:
our exercise holds debt and backing fixed, uses a common matching rate,
and characterizes its effect on the adoption thresholds. \cite{HuZhang2025} give a different reason why institutions may not use a public facility: from 2008 to 2010 banks avoided the discount window but borrowed through the Term Auction Facility, which had the same requirements, because the two carried different stigma. Here the obstacle is a small user network, which the matching transfer addresses directly.

We use a coordination model to ask a macroeconomic question: when does public debt supply the liquidity service that this literature assigns to it? The answer is two ranges of debt. Inside the wider range a positive network can exist, and at its edge the convenience value is discontinuous. Inside the narrower range that network is the long-run outcome. A temporary transfer can start a network that then sustains itself. The analysis does not replace models of optimal debt. It gives the conditions under which the liquidity service they assume is supplied.

Section~\ref{sec:environment} sets out the environment.
Section~\ref{sec:equilibria} characterizes the equilibria and their
potential. Section~\ref{sec:selection} introduces the logit rule, applies
it first to continuum renewal and then to finite-population stationary
selection, and characterizes the selection threshold.
Section~\ref{sec:policy} studies debt corridors, corrective transfers,
and convenience value. All proofs are in Appendix~\ref{app:proofs}; a
separate finite-horizon consistency result connects the two population
descriptions.

\section{Environment}
\label{sec:environment}

\paragraph{Debt and benchmark quality.}
The real face value of the eligible government claim is $x>0$, and
$\varphi>0$ measures the real resources backing those claims. Each unit
promises a real payment normalized to one. Let $s(x;\varphi)\in(0,1]$
denote its \emph{payoff retention}: the fraction of this unit payment
delivered to its holder.\footnote{Retention below one can come from partial default or from inflation. The parameter $\varphi$ is the fiscal backing of the claim. \cite{BassettoCui2018} study how that backing works when interest rates are low.} Let $m(x)>0$ measure effective market depth;
its scale incorporates infrastructure and legal usability. Define
\emph{benchmark quality}, also referred to as \emph{benchmark strength}, by
\begin{equation}
 b(x;\varphi)=m(x)s(x;\varphi).
 \label{eq:strength}
\end{equation}
Quality combines ease of use with the value retained by the claim. The two factors follow the distinction between an asset's liquidity and its safety in \cite{GeromichalosHerrenbrueckLee2023}.
The adoption analysis takes these functions as given;
Section~\ref{sec:policy} specifies how they vary with eligible debt supply.

\paragraph{Institutions.}

There is a unit mass of institutions. Each chooses $a_i\in\{0,1\}$,
where adoption means operating with the government claim as its
reference; the outside convention yields a payoff normalized to zero. Costs have cross-sectional
distribution $F$. An institution's $c_i$ is the recurring cost of maintaining the
operational capacity to use the claim as its reference, relative to
the outside convention. Examples include operating valuation and
settlement systems and retaining staff or external services for
collateral monitoring, margin management, and reconciliation.
These activities continue after initial setup.\footnote{
For example, \cite{SIX2022Collateral} describes ongoing collateral
transfers and multiple daily valuation checks in its collateral
management service. The fee schedule in \cite{LCH2025RepoClearFees}
distinguishes minimum monthly charges from one-time onboarding fees.}
We model the reference-specific maintenance burden as a cost of keeping
this capability active, rather than a fee for each transaction.


Let $A=\int a_i\dd i$ denote aggregate adoption. An adopter encounters
compatible counterparties at rate $m(x)A$ and obtains a benefit scaled
by payoff retention $s(x;\varphi)$. Its gross benefit is therefore
$b(x;\varphi)A$. This is a thick-market externality of the kind in \cite{Diamond1982}: the return to participating rises with the number of other participants. Writing $b=b(x;\varphi)$, the payoff difference of
adoption is
\begin{equation}
 u_i(1;A)-u_i(0;A)=bA-c_i.
 \label{eq:payoff}
\end{equation}
The adoption analysis initially holds $b>0$ fixed.

\begin{assumption}[Single-peaked cost distribution]
\label{ass:costs}
The distribution has support $[0,\infty)$, $F(0)=0$, and
$\lim_{c\to\infty}F(c)=1$. Its density $f$ is positive and continuously
differentiable on $(0,\infty)$, extends continuously to zero, and
$f(0)=0$. There is a mode $c_m>0$ such that
\begin{equation}
 f'(c)>0\quad\text{for }0<c<c_m,\qquad
 f'(c)<0\quad\text{for }c>c_m.
 \label{eq:singlepeak}
\end{equation}
\end{assumption}

Single-peakedness places a shape restriction on the cost density without
choosing a parametric family. Common distributions satisfying
Assumption~\ref{ass:costs} include the log-normal and, when the shape parameter exceeds one, the Gamma, Weibull, and log-logistic distributions. In appendix \ref{sec: density extension}, we relax condition $f(0)=0$ to allow for more cost distributions. We show that most results would go through.

Write $F^{-1}$ for the inverse cost distribution, with $F^{-1}(0)=0$.
For a share $A$, the value $F^{-1}(A)$ is the cost of the marginal
institution when adoption is assigned to the lowest-cost users. Define
\begin{equation}
 \Ccal(A)=\int_0^A F^{-1}(u)\dd u
          =\int_0^{F^{-1}(A)}c\dd F(c),\qquad 0\leq A<1.
 \label{eq:cost}
\end{equation}
This is the lowest aggregate implementation cost of an adoption share
$A$, obtained by assigning adoption to the lowest-cost institutions.
At $A=1$, use the extended value
$\Ccal(1)=\lim_{A\uparrow1}\Ccal(A)$, possibly $+\infty$.

\begin{definition}[Adoption equilibrium]
\label{def:equilibrium}
An adoption equilibrium at strength $b$ is a share $A\in[0,1]$ satisfying
\begin{equation}
 A=F(bA).
 \label{eq:equilibrium}
\end{equation}
Institutions with $c<bA$ adopt and those with $c>bA$ do not; the indifferent
type has zero mass.
\end{definition}

\section{Adoption equilibria and potential}
\label{sec:equilibria}

Zero adoption is always an equilibrium. If no institution uses the
government claim as its reference, an individual institution obtains no
network benefit from adopting but still incurs its implementation cost,
so it has no incentive to adopt on its own. The more interesting question
is whether the model also admits nontrivial equilibria with positive
adoption. When do such equilibria exist, and if they exist, how many can
there be?

To answer these questions, consider a positive equilibrium with a marginal
adopter of cost $c$. Since the marginal adopter satisfies $c=bA$ and the
adoption share is $A=F(c)$, the benchmark strength needed to sustain a
cutoff $c$ is $\frac{c}{F(c)}$.
Define the lowest such strength by
\begin{equation}
 b_f\equiv\min_{c>0}\frac{c}{F(c)},\qquad
 c_f\equiv\argmin_{c>0}\frac{c}{F(c)},\qquad
 A_f\equiv F(c_f).
 \label{eq:fold}
\end{equation}
The next theorem shows that, under Assumption~\ref{ass:costs}, these
objects are well defined and that $b_f$ completely characterizes the
existence and multiplicity of positive-adoption equilibria.

\begin{theorem}[Equilibrium existence and multiplicity]
\label{thm:equilibria}
Under Assumption~\ref{ass:costs}, strictly positive equilibria exist if
and only if $b\geq b_f$. At $b=b_f$, the unique positive equilibrium
is $A_f$. If $b>b_f$, there are exactly two positive equilibria
$A_-(b)<A_f<A_+(b)$.
\end{theorem}

The existence threshold measures how useful the reference must be before
any group of institutions can sustain common use. A small group offers
few compatible counterparties, so even low-cost institutions have little
reason to join. A larger group offers more benefits, but sustaining it
requires participation by institutions with higher implementation costs.
Below the threshold, no positive network size balances these two forces.

Above the threshold, the feedback from common use is strong enough to
support both a smaller and a larger network. In each equilibrium, the
benefit available at that network size covers the costs of precisely
those institutions using it. Neither equilibrium removes the alternative
of staying out together. The theorem therefore establishes the
possibility of adoption, rather than predicting that institutions must
coordinate on it.

\paragraph{Potential geometry.}
For a cutoff allocation with adoption share $A$, define the potential
\begin{equation}
 \Pcal(A;b)=\frac b2 A^2-\Ccal(A),\qquad 0\leq A\leq1.
 \label{eq:potential}
\end{equation}
For a given adoption share, assigning adoption to the lowest-cost
institutions minimizes implementation costs. 
The derivative has a direct incentive interpretation:
\begin{equation}
 \Pcal_A(A;b)=bA-F^{-1}(A).
 \label{eq:potentialderivatives}
\end{equation}
Increasing the cutoff slightly brings in an institution with cost $F^{-1}(A)$.
The derivative is its private adoption gain at the prevailing
network size. Thus the positive stationary points of $\Pcal$, i.e., $\Pcal_A(A;b)=0$, are exactly
the positive adoption equilibria.

Zero is a strict one-sided local potential maximum for every finite $b$:
near zero, the marginal cost $F^{-1}(A)$ exceeds $bA$. When $b>b_f$, the
lower root $A_-$ is a strict local minimum and the upper root $A_+$ is
a strict local maximum. At $b=b_f$, the two positive stationary points merge at $A_f$ and
disappear when $b$ falls further. This meeting point is the \emph{fold}.
The potential is decreasing on either side of it, so the fold itself is
neither a local maximum nor a local minimum. When
$b\leq b_f$, zero is the unique global potential maximizer. (See the proof of Theorem~\ref{thm:equilibria} for a formal argument.)

\begin{figure}[!htbp]
\centering
\begin{tikzpicture}[x=1cm,y=1cm,benchmark text]
\begin{scope}[shift={(.73,0)}]
  \def\plotbmin{1.78}
  \def\plotbmax{3.02}
  \def\plotbp{2.17506296272756}
  \def\plotwidth{5.86}
  \pgfmathsetmacro{\foldx}{\plotwidth*(2-\plotbmin)/(\plotbmax-\plotbmin)}
  \pgfmathsetmacro{\selectx}{\plotwidth*(\plotbp-\plotbmin)/(\plotbmax-\plotbmin)}
  \pgfmathsetmacro{\tmax}{sqrt(1-4/(\plotbmax*\plotbmax))}
  \node[benchmark title] at (2.93,4.78) {(a) Existence and local stability};
  \draw[benchmark axis] (0,0) -- (6.03,0);
  \draw[benchmark axis] (0,0) -- (0,4.43);
  \draw[benchmark guide] (\foldx,.02) -- (\foldx,4.2);
  \draw[draw=benchmarkteal,dash pattern=on 4pt off 1.4pt on .8pt off 1.4pt,line width=.65pt]
    (\selectx,.02) -- (\selectx,4.2);
  \draw[benchmark critical] (0,.17) -- (\plotwidth,.17);
  \draw[benchmark high] plot[variable=\t,domain=0:\tmax,samples=160]
    ({\plotwidth*(2/sqrt(1-\t*\t)-\plotbmin)/(\plotbmax-\plotbmin)},
     {.17+4.06*(1+\t)/2});
  \draw[benchmark low] plot[variable=\t,domain=0:\tmax,samples=160]
    ({\plotwidth*(2/sqrt(1-\t*\t)-\plotbmin)/(\plotbmax-\plotbmin)},
     {.17+4.06*(1-\t)/2});
  \draw[benchmark guide] (0,2.20) -- (\foldx,2.20);
  \draw[draw=benchmarkblue,fill=white,line width=.85pt] (\foldx,2.20) circle[radius=2.4pt];
  \node[anchor=east] at (-.11,.17) {$0$};
  \node[anchor=east] at (-.11,2.20) {$A_f$};
  \node[anchor=north] at (\foldx,-.09) {$b_f$};
  \node[anchor=north,text=benchmarkteal] at (\selectx,-.09) {$b_p$};
  \node at (2.93,-.88) {Benchmark quality $b$};
  \node[rotate=90] at (-.94,2.22) {Adoption share $A$};
  \draw[benchmark high] (2.38,2.30) -- (2.96,2.30);
  \node[anchor=west] at (3.04,2.30) {$A_+$: locally stable};
  \draw[benchmark low] (2.38,1.90) -- (2.96,1.90);
  \node[anchor=west] at (3.04,1.90) {$A_-$: critical mass};
  \draw[benchmark critical] (2.38,1.50) -- (2.96,1.50);
  \node[anchor=west] at (3.04,1.50) {$A=0$: locally stable};
\end{scope}
\begin{scope}[shift={(8.28,0)}]
  \def\plotwidth{5.78}
  \def\Aend{.9}
  \def\Pmin{-.215}
  \def\Prange{.295}
  \def\plotheight{4.22}
  \def\bhigh{2.4}
  \pgfmathsetmacro{\Aminus}{(1-sqrt(1-4/(\bhigh*\bhigh)))/2}
  \pgfmathsetmacro{\Aplus}{1-\Aminus}
  \pgfmathsetmacro{\Yzero}{\plotheight*(0-\Pmin)/\Prange}
  \pgfmathsetmacro{\Yfold}{\plotheight*(benchmarkP(.5,2)-\Pmin)/\Prange}
  \pgfmathsetmacro{\Xfold}{\plotwidth*.5/\Aend}
  \pgfmathsetmacro{\Xminus}{\plotwidth*\Aminus/\Aend}
  \pgfmathsetmacro{\Xplus}{\plotwidth*\Aplus/\Aend}
  \pgfmathsetmacro{\Yminus}{\plotheight*(benchmarkP(\Aminus,\bhigh)-\Pmin)/\Prange}
  \pgfmathsetmacro{\Yplus}{\plotheight*(benchmarkP(\Aplus,\bhigh)-\Pmin)/\Prange}
  \node[benchmark title] at (2.89,4.78) {(b) Potential around $b_f$};
  \draw[benchmark axis] (0,0) -- (5.97,0);
  \draw[benchmark axis] (0,0) -- (0,4.43);
  \draw[draw=benchmarkgray,line width=.45pt] (0,\Yzero) -- (\plotwidth,\Yzero);
  \node[anchor=east] at (-.10,\Yzero) {$0$};
  \node[rotate=90] at (-.57,2.22) {Potential $\Pcal$};
  \draw[benchmark guide] (\Xminus,0) -- (\Xminus,\Yminus);
  \draw[benchmark guide] (\Xfold,0) -- (\Xfold,\Yfold);
  \draw[benchmark guide] (\Xplus,0) -- (\Xplus,\Yplus);
  \draw[benchmark high] plot[variable=\a,domain=0:\Aend,samples=301]
    ({\plotwidth*\a/\Aend},{\plotheight*(benchmarkP(\a,\bhigh)-\Pmin)/\Prange});
  \draw[benchmark critical] plot[variable=\a,domain=0:.5,samples=151]
    ({\plotwidth*\a/\Aend},{\plotheight*(benchmarkP(\a,2)-\Pmin)/\Prange});
  \draw[benchmark critical] plot[variable=\a,domain=.5:\Aend,samples=151]
    ({\plotwidth*\a/\Aend},{\plotheight*(benchmarkP(\a,2)-\Pmin)/\Prange});
  \draw[benchmark low] plot[variable=\a,domain=0:\Aend,samples=301]
    ({\plotwidth*\a/\Aend},{\plotheight*(benchmarkP(\a,1.85)-\Pmin)/\Prange});
  \draw[draw=benchmarkpurple!65,line width=.4pt] ({\Xfold-.28},\Yfold) -- ({\Xfold+.28},\Yfold);
  \fill[benchmarkblue] (\Xminus,\Yminus) circle[radius=1.9pt];
  \fill[benchmarkblue] (\Xplus,\Yplus) circle[radius=1.9pt];
  \draw[draw=benchmarkpurple,fill=white,line width=.8pt] (\Xfold,\Yfold) circle[radius=2.1pt];
  \node[anchor=north] at (\Xminus,-.10) {$A_-(b_1)$};
  \node[anchor=north] at (\Xfold,-.10) {$A_f$};
  \node[anchor=north] at (\Xplus,-.10) {$A_+(b_1)$};
  \node at (2.89,-.88) {Adoption share $A$};
  \draw[benchmark high] (.34,4.08) -- (.91,4.08);
  \node[anchor=west] at (1.01,4.08) {$b_1>b_f$};
  \draw[benchmark critical] (.34,3.71) -- (.91,3.71);
  \node[anchor=west] at (1.01,3.71) {$b_2=b_f$};
  \draw[benchmark low] (.34,3.34) -- (.91,3.34);
  \node[anchor=west] at (1.01,3.34) {$b_3<b_f$};
\end{scope}
\end{tikzpicture}
\caption{Equilibrium branches and potential geometry.
Panel (a) shows the positive branches meeting at $b_f$ and previews the
selection threshold $b_p$. The stability labels concern the branches
away from the fold. In panel (b), $b_1>b_f$ gives a local minimum and a
local maximum. At $b_2=b_f$, the potential remains decreasing, with a
single positive stationary point at $A_f$ below zero. For $b_3<b_f$,
there is no positive stationary point.  The diagrams are qualitative; the comparison values $b_i$ are
specific to this figure.}
\label{fig:existence-potential}
\end{figure}
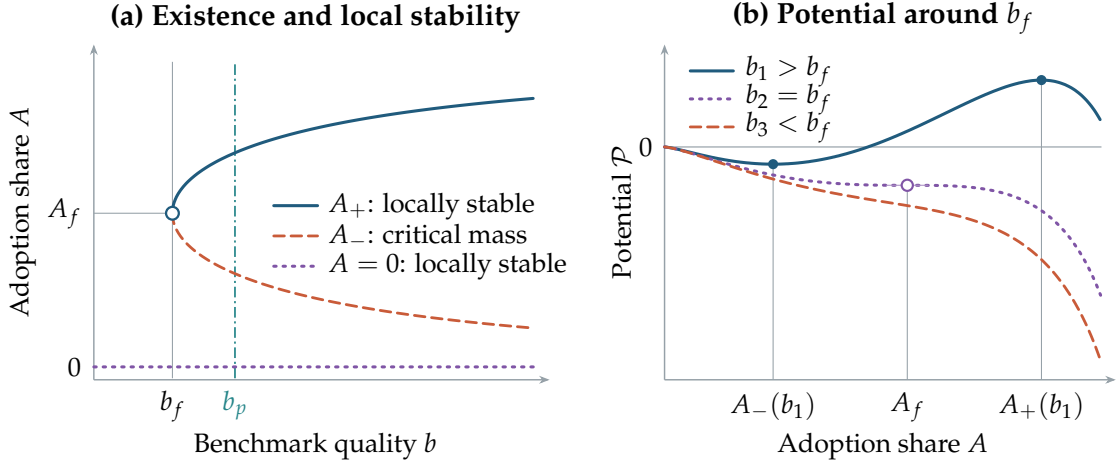

Figure~\ref{fig:existence-potential} illustrates the equilibrium branches
and their potential geometry. Panel (b) distinguishes two positive
stationary points above $b_f$ from one at $b_f$ and none below it. At the
threshold, the potential has a horizontal tangent at $A_f$ but continues
to decrease on both sides; this point lies strictly below the value at
nonadoption. The local stability labels in panel (a) are established in
Proposition~\ref{prop:dynamics}. The marker $b_p$ previews the selection
threshold characterized in Theorem~\ref{thm:thresholds}.

We define the set of potential maximizers
\begin{equation}
 \Mcal(b)=\argmax_{0\leq A\leq1}\Pcal(A;b).
 \label{eq:maximizers}
\end{equation}
This set is nonempty and compact. 
For $b>b_f$, the only global candidates are $0$ and $A_+(b)$.
Local geometry alone does not rank these two candidates. The next section
first explains why their global potential ranking matters for behavior,
then pins down the potential maximizer.

\section{Adjustment and long-run selection}
\label{sec:selection}

The equilibrium equation describes which adoption shares are
self-consistent, but it does not describe how institutions revise their
decisions. We first introduce the individual logit choice rule. We then
apply it to the continuum of institutions to study renewal from a given
initial user base. Finally, we introduce a finite population to study the
stationary distribution of noisy revisions. The same payoff comparison
and revision opportunities govern both population descriptions.
Benchmark strength $b$ and implementation costs remain fixed throughout
this section.


Each institution receives revision opportunities according to an
independent Poisson clock with rate one. At an opportunity, it compares
the current payoff from adoption with the payoff from nonadoption,
taking the other institutions' current choices as given. Its chosen
action remains in place until its next revision opportunity.

Let $g$ denote the payoff gain from adoption rather than nonadoption.
The institution chooses adoption with probability
\begin{equation}
 L(g;\eta)=\frac{\exp(g/\eta)}{1+\exp(g/\eta)},\qquad \eta>0.
 \label{eq:logit}
\end{equation}
It chooses nonadoption with the complementary probability. The parameter $\eta$
measures revision noise. The more profitable action is more likely, but
both actions have positive probability at any finite payoff difference.
A revision can therefore change an institution's action or leave it
unchanged.

As $\eta$ vanishes, the rule approaches best response: adoption is chosen
when $g>0$, nonadoption when $g<0$, and the two actions have equal
probability at exact indifference. For a cost-$c$ institution in the
continuum economy, the payoff gain is $g=bA-c$. 

\subsection{Continuum renewal and critical mass}
\label{sec:renewal}

 Fix $\eta>0$. Let $a(c,t;\eta)$
be the fraction of cost-$c$ institutions that adopt at time $t$, and let
\[
 A(t;\eta)=\int_0^\infty a(c,t;\eta)\dd F(c)
\]
be aggregate adoption. Both quantities are scalar shares; the parameter
after the semicolon records revision noise. Take an initial allocation
with aggregate share $A(0;\eta)=A_0\in[0,1]$.

A cost-$c$ institution adopts with probability $L(bA(t;\eta)-c;\eta)$
whenever it revises. At unit-rate revision, the inflow from current
nonadopters is
\[
 [1-a(c,t;\eta)]L(bA(t;\eta)-c;\eta),
\]
while the outflow from current adopters is
\[
 a(c,t;\eta)[1-L(bA(t;\eta)-c;\eta)].
\]
Subtracting outflows from inflows gives
\[
 \partial_t a(c,t;\eta)
 =L(bA(t;\eta)-c;\eta)-a(c,t;\eta).
\]
Integrating these type-level flows over costs yields
\begin{equation}
 \dot A(t;\eta)
 =\int_0^\infty L(bA(t;\eta)-c;\eta)\dd F(c)-A(t;\eta).
 \label{eq:noisyrenewal}
\end{equation}
The integral is the share that would choose adoption at a revision
opportunity, given the current network. Actual adoption moves toward
that share as institutions reconsider their decisions. In this
continuum description, probabilistic individual choices enter through
deterministic adoption flows, so aggregate adoption follows a
deterministic path from its initial share.\footnote{The derivation does not
require the current adopters to be the lowest-cost institutions.}

As $\eta\downarrow0$, institutions choose the more profitable action.
Since $F$ is continuous, the cost type exactly indifferent at a given
adoption share has zero mass. The aggregate best response therefore
converges to $F(bA)$. Writing $A(t)$ for noiseless adoption gives
\begin{equation}
 \dot A(t)=F(bA(t))-A(t),\qquad A(0)=A_0\in[0,1].
 \label{eq:renewal}
\end{equation}
Equivalently, the noiseless type fractions satisfy
$\partial_t a(c,t)=\mathbf{1}\{c<bA(t)\}-a(c,t)$ almost everywhere.
The revision opportunities and payoff comparison are unchanged; only
the noise in the choice at an opportunity has disappeared. We now ask
where this renewal process leads from a given initial installed base. Note that equilibria are exactly the fixed points of this process.

\begin{proposition}[Noiseless adjustment dynamics]
\label{prop:dynamics}
Under Assumption~\ref{ass:costs}, zero is locally asymptotically stable
under \eqref{eq:renewal} for every finite $b$.
If $b<b_f$, every initial share converges to zero. If $b>b_f$,
\[
 \begin{split}
 A_0\in[0,A_-(b))&\ \Longrightarrow\ A(t)\longrightarrow0,\\
 A_0\in(A_-(b),1]&\ \Longrightarrow\ A(t)\longrightarrow A_+(b).
 \end{split}
\]
The lower root is unstable.
At $b=b_f$, initial shares below $A_f$ converge to zero and those at or
above $A_f$ converge to $A_f$.

\end{proposition}

The lower equilibrium is a critical mass because an institution's reason
to use the reference depends on how many counterparties already use it.
Below that share, the available benefits attract fewer users than the
network currently has, so adoption declines and reinforces further exit.
Above it, the network moves toward the high-adoption equilibrium.
The same benchmark quality can therefore sustain an established user
base while failing to support a smaller one. This dependence on the
installed base follows from the adoption incentives, even though
institutions face no separate switching cost. These basins explain the
stability labels in Figure~\ref{fig:existence-potential}(a).

\begin{corollary}[Launching from an initial share]
\label{cor:launch}
For $A_0\in(0,A_f)$, convergence under \eqref{eq:renewal} to the high
equilibrium occurs if and only if
\begin{equation}
 b>b_{\mathrm{initial\ share}}(A_0)
   \equiv\frac{F^{-1}(A_0)}{A_0}>b_f.
 \label{eq:initialshare}
\end{equation}
The strength threshold decreases strictly with $A_0$ and diverges as
$A_0\downarrow0$.
\end{corollary}

Launching a reference and maintaining one impose different requirements.
A small initial group offers few compatible counterparties, so the claim
must be especially useful to attract enough additional institutions.
A larger initial group supplies more of that benefit itself and can grow
at a lower benchmark strength. The corollary states the precise trade-off
between initial participation and the quality needed for eventual
convergence to the high-adoption equilibrium.

In particular, the existence of a positive equilibrium does not guarantee
that an arbitrarily small launch will reach it. With no revision noise,
zero adoption cannot restart on its own, and even a small positive base
may continue to contract. These deterministic basins describe adjustment
from a given initial share. To study long-run selection when individual
choices remain noisy, we next apply the same revision rule to a finite
population.

\subsection{Finite-population revision and stationary selection}
\label{sec:finite}
\label{sec:stationary}

Stationary selection concerns the long-run frequencies of adoption
configurations when institutions continue to revise with positive noise.
For this question, we now introduce a finite population. This preserves
random fluctuations in aggregate adoption and makes the process an
irreducible Markov chain.

Consider a finite population setting with $n$ institutions. For $n\geq2$, let
\[
 \mathbf{a}=(a_1,\ldots,a_n)\in\{0,1\}^n
\]
be the profile of all institutions' adoption actions. The vector
$\mathbf{a}_{-i}$ collects the actions other than $i$'s. Individual actions
$a_i$ remain scalar. Choose a representative, deterministic cost array
and define aggregate adoption by
\begin{equation}
 c_{i,n}=F^{-1}\!\left(\frac{i}{n+1}\right),\quad i=1,\ldots,n,
 \qquad A_n(\mathbf{a})=\frac1n\sum_{i=1}^n a_i.
 \label{eq:finitecosts}
\end{equation}
The empirical cost distribution converges to $F$ as $n$ grows; costs do
not change at revision dates. Institution $i$'s gain from choosing adoption
rather than nonadoption is
\begin{equation}
 g_{i,n}(\mathbf{a}_{-i})
 =\frac bn\sum_{j\ne i}a_j-c_{i,n}.
 \label{eq:finitegain}
\end{equation}
Excluding its own action makes the comparison depend only on the other
institutions' choices. Its difference from $bA_n-c_{i,n}$ is at most
$b/n$ and disappears in the large-population limit.

\paragraph{The revision process and its potential.}
Each institution uses the unit-rate Poisson revision clock and logit rule \ref{eq:logit}. At a revision opportunity, institution $i$
chooses adoption with probability $L(g_{i,n}(\mathbf{a}_{-i});\eta)$ and
nonadoption with the complementary probability. Revision clocks and
choice draws are independent across institutions conditional on the
current profile. The state records the individual actions, not only the
aggregate share, because institutions have different fixed costs.

The revision process defines a continuous-time Markov chain on the
set of adoption profiles $\{0,1\}^n$.
Total revision opportunities arrive at rate $n$; conditional on an
opportunity, the revising institution is uniform over the population.
The chain observed at those opportunities is therefore the usual
one-at-a-time logit revision chain. The continuous-time process and this
embedded chain have the same invariant distribution.

The finite game has the exact potential
\begin{equation}
 \Phi_n(\mathbf{a};b)
 =\frac bn\sum_{i<j}a_i a_j-\sum_{i=1}^n c_{i,n}a_i.
 \label{eq:finitepotential}
\end{equation}
When one institution switches from nonadoption to adoption, the increase
in $\Phi_n$ is exactly its payoff gain in \eqref{eq:finitegain}.
For a fixed number of adopters, the first term is unchanged across
profiles, so the highest potential is obtained by selecting the
lowest-cost institutions. Normalizing by $n$ and increasing the population
then gives the cutoff potential $\Pcal(A;b)$ in \eqref{eq:potential}.
The approximation is established in the proof of
Theorem~\ref{thm:logit}.

\paragraph{The invariant distribution.}
For fixed $n$ and $\eta>0$, every profile can be reached through a
sequence of revisions with positive probability. The finite
continuous-time chain is therefore irreducible and has a unique
invariant distribution. This distribution describes the long-run
frequencies of adoption profiles.

The invariant distribution can be obtained directly from the logit
revision rule. Let $q_{n,\eta}(\mathbf{a},\mathbf{a}')$ denote the
transition rate from profile $\mathbf{a}$ to a distinct profile
$\mathbf{a}'$. A sufficient condition for a probability distribution
$\mu$ to be invariant is \emph{detailed balance}:
\[
 \mu(\mathbf{a})q_{n,\eta}(\mathbf{a},\mathbf{a}')
 =
 \mu(\mathbf{a}')q_{n,\eta}(\mathbf{a}',\mathbf{a})
 \qquad\text{for all distinct }\mathbf{a},\mathbf{a}'.
\]
The left-hand side is the probability flow from $\mathbf{a}$ to
$\mathbf{a}'$, and the right-hand side is the reverse flow. Detailed
balance equates these flows pair by pair; summing over the other
profiles then equates the total inflow and outflow at each profile,
so the distribution remains unchanged over time.

To verify this condition, consider neighboring profiles
$\mathbf{a}$ and $\mathbf{a}'$ that differ only because institution
$i$ switches from nonadoption to adoption. The reverse revision
faces the same actions $\mathbf{a}_{-i}$ of the other institutions.
Since the individual revision rate is one, the logit rule gives
\[
 \frac{q_{n,\eta}(\mathbf{a},\mathbf{a}')}
      {q_{n,\eta}(\mathbf{a}',\mathbf{a})}
 =
 \frac{L(g_{i,n}(\mathbf{a}_{-i});\eta)}
      {1-L(g_{i,n}(\mathbf{a}_{-i});\eta)}
 =
 \exp\!\left\{
 \frac{\Phi_n(\mathbf{a}';b)-\Phi_n(\mathbf{a};b)}{\eta}
 \right\}.
\]
The last equality uses the exact-potential property: the change in
$\Phi_n$ equals the revising institution's payoff gain. Thus assigning
probabilities proportional to $\exp\{\Phi_n/\eta\}$ satisfies detailed
balance. Profiles that differ in more than one action have zero
direct transition rates in both directions. Consequently, the unique
invariant distribution is
\begin{equation}
 \mu_{n,\eta}(\mathbf{a})
 =
 \frac{\exp\{\Phi_n(\mathbf{a};b)/\eta\}}
 {\displaystyle\sum_{\widetilde{\mathbf{a}}\in\{0,1\}^n}
  \exp\{\Phi_n(\widetilde{\mathbf{a}};b)/\eta\}}.
 \label{eq:gibbs}
\end{equation}

The next theorem establishes which aggregate adoption levels this
distribution selects as the population grows and revision noise
vanishes. For a nonempty set $S\subseteq[0,1]$, write
$\dist(A,S)=\inf_{y\in S}|A-y|$.

\begin{theorem}[Stationary logit selection]
\label{thm:logit}
Suppose Assumption~\ref{ass:costs} holds and fix $b>0$.
For every sequence $\eta_n>0$ with $\eta_n\to0$ and every $\delta>0$,
\begin{equation}
 \mu_{n,\eta_n}\!\left(
 \dist(A_n(\mathbf{a}),\Mcal(b))\geq\delta
 \right)\longrightarrow0
 \quad\text{as }n\to\infty.
 \label{eq:concentration}
\end{equation}
Thus stationary adoption concentrates on the set of global potential
maximizers. In particular, whenever $\Mcal(b)=\{A^*(b)\}$, stationary
adoption converges in probability to the unique potential maximizer:
\[
 A_n(\mathbf{a})\xrightarrow{\mathbb P}A^*(b),
 \qquad \mathbf{a}\sim\mu_{n,\eta_n}.
\]
\end{theorem}

The theorem explains why individual revisions can select one equilibrium
without any institution choosing the outcome for the network as a whole.
Each revising institution responds only to its own gain. The exact-potential
property translates those gains into the relative probability flows
between adoption profiles, and the Gibbs measure ranks their long-run
frequencies accordingly. As noise becomes small and the population grows,
stationary adoption is concentrated near the global potential maximizers.
When there is only one such maximizer, almost all stationary probability
is assigned to adoption shares near that outcome; when several tie,
the theorem does not choose their limiting weights.

This criterion is stronger than resistance to a small loss of users.
A high-adoption network can be locally persistent and still lose the
long-run comparison with nonadoption. Nor does the criterion simply favor
the network with more participants: the costs of those participants enter
the potential as well. The proof accounts for all possible assignments of
adoption across institutions, including profiles in which some high-cost
institutions adopt and some low-cost institutions do not. It does not
assume that revisions remain within cost-ordered profiles.

The same proof also yields convergence of expected normalized
potential:
\[
 \E_{\mu_{n,\eta_n}}
 \!\left[\frac{\Phi_n(\mathbf{a};b)}{n}\right]
 \longrightarrow
 \max_{0\leq A\leq1}\Pcal(A;b).
\]
These are stationary selection results: the long-run distribution
is taken at each finite $n$ and positive $\eta$ before the
large-population, vanishing-noise limit. They do not assert that a
network starting from a particular installed base immediately moves
to the potential-maximizing equilibrium.

\paragraph{Why the order of limits matters.}
Section~\ref{sec:renewal} starts with a continuum, derives its adoption
flows, and studies the noiseless path as time passes. The finite model
introduced here also provides a precise connection to that description.
Lemma~\ref{lem:finitehorizon} in Appendix~\ref{app:finitehorizon} shows
that, on any fixed interval $[0,T]$, its aggregate path approaches
\eqref{eq:noisyrenewal} as the population grows at fixed positive noise.
If the population grows and noise vanishes jointly, the limiting path is
\eqref{eq:renewal}, provided initial aggregate adoption converges to
$A_0$. The initial adopters need not be ordered by cost.

This gives a finite-population interpretation of deterministic
adjustment. First take a large population and vanishing noise while
keeping the observation horizon $T$ fixed. The resulting approximation
removes aggregate fluctuations and gives the noiseless renewal equation.
Only then study the solution of that equation as $t\to\infty$.
Proposition~\ref{prop:dynamics} shows that its destination depends on
the initial installed base. A close approximation on each fixed time
interval does not guarantee closeness over arbitrarily long periods at
a given finite population size and positive noise level.

Stationary selection takes time to infinity first. Hold $n$ finite and
$\eta>0$, and let the distribution of the action profile converge to
$\mu_{n,\eta}$. Only afterward let the population grow and noise vanish,
as in Theorem~\ref{thm:logit}. In this order, the finite system has time
to move between adoption configurations before its fluctuations
disappear. The limiting distribution therefore need not select the same
outcome as noiseless adjustment from a particular initial share.

Keeping $\eta>0$ in the continuum equation does not reverse this order.
That equation already replaces individual random choices by aggregate
flows. Given an initial share, its path is deterministic, and a path at
a steady state has no random aggregate disturbance to move it away.
Following that equation for arbitrarily long therefore does not compute
the invariant distribution of the finite system. Local persistence and
stationary selection concern different long-run comparisons under the
same individual revision rule.

\subsection{The equilibrium-selection threshold}
\label{sec:thresholds}

The revision mechanism provides a reason to compare global potentials.
It remains to determine when positive adoption wins that comparison.
Zero has potential zero. For $b>b_f$, the only competing global candidate
is $A_+(b)$, so selection reduces to the sign of
$\Pcal(A_+(b);b)$.

\begin{theorem}[Equilibrium selection]
\label{thm:thresholds}
Under Assumption~\ref{ass:costs}, there is a unique finite threshold $b_p$
and a positive share $A_p=A_+(b_p)$ such that
\begin{equation}
 \Mcal(b)=
 \begin{cases}
  \{0\},&b<b_p,\\
  \{0,A_p\},&b=b_p,\\
  \{A_+(b)\},&b>b_p.
 \end{cases}
 \label{eq:selection}
\end{equation}
The threshold is characterized by
\begin{equation}
 b_p=\min_{0<A<1}\frac{2\Ccal(A)}{A^2},
 \label{eq:bp}
\end{equation}
with the minimum attained uniquely at $A_p$.
\end{theorem}

\begin{figure}[!htbp]
\centering
\begin{tikzpicture}[x=1cm,y=1cm,benchmark text]
\begin{scope}[shift={(.85,0)}]
  \def\plotwidth{10.6}
  \def\Aend{.9}
  \def\Pmin{-.135}
  \def\Prange{.22}
  \def\plotheight{5.35}
  \def\plotbp{2.17506296272756}
  \pgfmathsetmacro{\Ap}{(1+sqrt(1-4/(\plotbp*\plotbp)))/2}
  \pgfmathsetmacro{\Xp}{\plotwidth*\Ap/\Aend}
  \pgfmathsetmacro{\Yzero}{\plotheight*(0-\Pmin)/\Prange}
  \node[benchmark title] at (5.3,5.90) {Potential geometry};
  \draw[benchmark axis] (0,0) -- (10.82,0);
  \draw[benchmark axis] (0,0) -- (0,5.55);
  \draw[draw=benchmarkgray,line width=.55pt] (0,\Yzero) -- (\plotwidth,\Yzero);
  \node[anchor=east] at (-.13,\Yzero) {$0$};
  \node[rotate=90] at (-.68,2.78) {Potential $\Pcal$};
  \node[anchor=north] at (0,-.10) {$0$};
  \node at (5.3,-.84) {Adoption share $A$};
  \draw[benchmark guide] (\Xp,0) -- (\Xp,\Yzero);
  \draw[benchmark high] plot[variable=\a,domain=0:\Aend,samples=401]
    ({\plotwidth*\a/\Aend},{\plotheight*(benchmarkP(\a,2.4)-\Pmin)/\Prange});
  \draw[benchmark critical] plot[variable=\a,domain=0:\Ap,samples=251]
    ({\plotwidth*\a/\Aend},{\plotheight*(benchmarkP(\a,\plotbp)-\Pmin)/\Prange});
  \draw[benchmark critical] plot[variable=\a,domain=\Ap:\Aend,samples=151]
    ({\plotwidth*\a/\Aend},{\plotheight*(benchmarkP(\a,\plotbp)-\Pmin)/\Prange});
  \draw[benchmark low] plot[variable=\a,domain=0:\Aend,samples=401]
    ({\plotwidth*\a/\Aend},{\plotheight*(benchmarkP(\a,2.05)-\Pmin)/\Prange});
  \draw[draw=benchmarkpurple,fill=white,line width=.9pt] (\Xp,\Yzero) circle[radius=2.8pt];
  \node[anchor=north] at (\Xp,-.10) {$A_p$};
  \node[text=benchmarkpurple,anchor=west] at ({\Xp-1.00},{\Yzero+.62}) {Equal potential};
  \draw[draw=benchmarkpurple,line width=.65pt,-{Stealth[length=3.5pt]}]
    ({\Xp+.20},{\Yzero+.48}) -- ({\Xp+.055},{\Yzero+.065});
  \draw[benchmark high] (.45,5.02) -- (1.20,5.02);
  \node[anchor=west] at (1.32,5.02) {$b>b_p$};
  \draw[benchmark critical] (.45,4.63) -- (1.20,4.63);
  \node[anchor=west] at (1.32,4.63) {$b=b_p$};
  \draw[benchmark low] (.45,4.24) -- (1.20,4.24);
  \node[anchor=west] at (1.32,4.24) {$b<b_p$};
\end{scope}
\end{tikzpicture}
\caption{Potential comparison and stationary selection.
The positive local maximum is above zero for $b>b_p$, equal to zero
at $A_p$ for $b=b_p$, and below zero for the illustrated
$b_f<b<b_p$. Since nonadoption always has zero potential, these cases
select the high-adoption equilibrium, permit both $0$ and $A_p$, and
select nonadoption, respectively. }
\label{fig:selection-potential}
\end{figure}
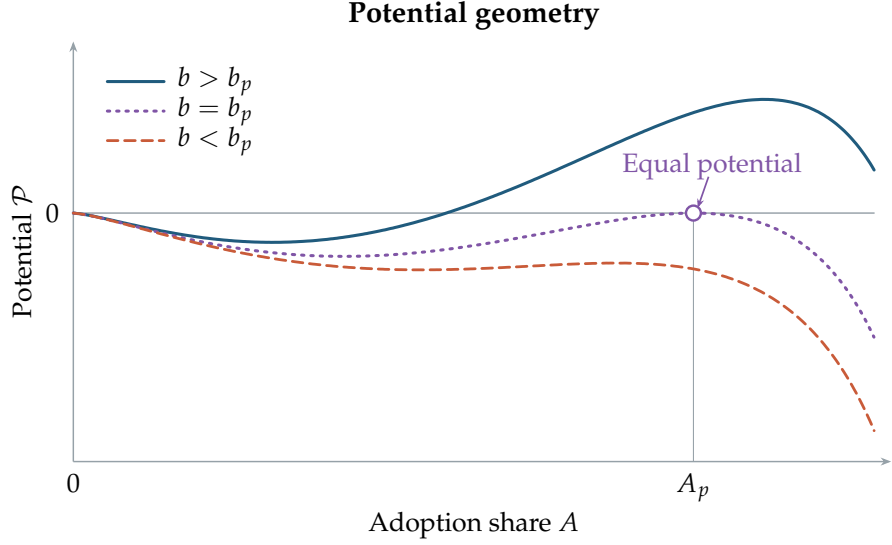

Combining Theorems~\ref{thm:logit} and~\ref{thm:thresholds}, in the
large-population, vanishing-noise limit, stationary adoption converges
in probability to zero for $b<b_p$ and to $A_+(b)$ for
$b>b_p$.\footnote{At $b=b_p$, stationary adoption concentrates on
$\{0,A_p\}$, but the result does not specify the limiting probability
assigned to each outcome.}
Figure~\ref{fig:selection-potential} shows the global comparison behind
this selection. The upper local maximum is above zero when the illustrated
strength exceeds $b_p$, touches zero at $b_p$, and lies below zero for strengths below $b_p$.  

Existence and selection therefore impose different requirements on
benchmark quality. The threshold $b_f$ is the lowest quality at which
a positive network can sustain itself. The threshold $b_p$ is the
quality at which the high-adoption equilibrium first matches
nonadoption in potential. These thresholds satisfy\footnote{The proof of
Theorem~\ref{thm:thresholds} establishes the bounds in
\eqref{eq:thresholdbounds}.}
\begin{equation}
 b_f<b_p<2b_f,\qquad A_p>A_f.
 \label{eq:thresholdbounds}
\end{equation}
Thus making a positive network feasible is not enough to make it the
long-run selected outcome. The bound holds for every distribution satisfying Assumption~\ref{ass:costs}: selection never requires more than twice the quality needed for existence.

The gap arises because the incentives supporting an operating network
are stronger than those supporting participation at smaller network
sizes. At $b_f$, the network with adoption share $A_f$ generates just
enough benefit to cover the cost of its marginal user. At every smaller
positive share, however, the network benefit falls short of the cost
of the marginal institution needed to sustain that share. The
potential records these private marginal incentives across network
sizes. Their negative values below $A_f$ leave the positive equilibrium
with lower potential than nonadoption at the point where it first
becomes feasible.

Higher quality strengthens the incentives to use the reference without
changing implementation costs. It therefore improves the potential
of adoption relative to nonadoption. The theorem shows that this
ranking reverses at a single threshold: at $b_p$, the high-adoption
equilibrium and nonadoption have equal potential; above $b_p$, the
high-adoption equilibrium wins the comparison. The interval
$(b_f,b_p)$ consequently consists of qualities at which a positive
network can sustain itself but is not selected by the stationary
logit criterion.


\FloatBarrier
\section{Debt corridors and corrective policy}
\label{sec:policy}

We now examine how the supply of eligible government debt affects its
usefulness as a common reference. Benchmark quality combines two
components. Effective market depth $m(x)$ captures how readily
institutions can obtain and trade the claim, given the infrastructure
that supports its use. We take it to be increasing in the face
value of eligible debt $x$: a larger supply makes the claim more
widely available and supports trading in it.
\cite{Fleming2002} finds that larger Treasury issues are more liquid.
Payoff retention
$s(x;\varphi)\in(0,1]$ is the fraction of the promised real payment
delivered to a holder. Holding the backing resources $\varphi$ fixed,
we take retention to be weakly decreasing in $x$, since additional
issuance spreads those resources across more promised payments.
A low interest rate relative to growth lowers the fiscal cost of that stock \citep{Blanchard2019,Reis2022}, but it does not undo the dilution.
Both components are continuous, and their product determines
benchmark quality:
\[
 b(x;\varphi)=m(x)s(x;\varphi).
\]

These two components create a trade-off between availability and
payoff dilution. When the debt stock is small relative to its backing,
retention can be close to one, but limited supply makes the claim
difficult to use as a common reference. Additional issuance then
primarily improves market depth. At high debt levels, the market is
already well supplied, so further issuance contributes little
additional depth. Retention, however, continues to decline as the
same backing supports more claims. The loss in retained payoff can
then outweigh the remaining gains from a deeper market.
The schedule is hump-shaped for a different reason from the debt Laffer curve in \cite{BrunnermeierMerkelSannikov2024} and the revenue-maximizing debt level in \cite{BayerBornLuetticke2023}. In those papers the liquidity value per unit falls as more debt is issued. Here depth rises with supply while backing per claim falls.

This trade-off motivates a quality schedule that first rises and
then falls with debt supply. A very small issue provides too little
depth, while a sufficiently large issue suffers enough payoff
dilution to undermine its usefulness. Rather than specify a
particular functional form for either component, we impose the
following shape restriction on their product.

\begin{assumption}[Single-peaked quality]
\label{ass:quality}
For each $\varphi>0$, $b(\cdot;\varphi)$ is positive and continuous
on $(0,\infty)$, tends to zero at both endpoints, and is strictly
increasing up to a unique $\widehat x(\varphi)$ and strictly
decreasing thereafter.
\end{assumption}

The peak $\widehat x(\varphi)$ separates the range in which the
benefits of greater depth dominate from the range in which payoff
dilution dominates. Comparing $b(x;\varphi)$ with the adoption thresholds
$b_f$ and $b_p$ therefore translates the existence and selection
requirements into restrictions on debt supply.
We do not solve the issuance problem of \cite{BassettoCui2024}, or the choice of maturity in \cite{JiangSargentWang2026}. The model has a single eligible claim, and we take its stock as given.

\subsection{Feasibility and coordination corridors}
\label{sec:corridors}

We now map the two private-adoption thresholds back into debt supply.
For any strength requirement $\tau>0$, define
\begin{equation}
 \Xcal(\tau)=\{x>0:b(x;\varphi)\geq\tau\},\qquad
 b_{\max}=b(\widehat x(\varphi);\varphi).
 \label{eq:corridordef}
\end{equation}
Write $\Xcal_f=\Xcal(b_f)$ for the feasibility corridor and
$\Xcal_p=\Xcal(b_p)$ for the coordination corridor. The first records
where a positive private equilibrium exists; the second records where a
positive equilibrium is a global potential maximizer. 

\begin{corollary}[Corridors and comparative statics]
\label{col:corridors}
Under Assumptions~\ref{ass:costs} and~\ref{ass:quality},  $\Xcal_p\subseteq\Xcal_f$. If $b_{\max}>b_p$,
both corridors are nondegenerate and the inclusion is strict. If
$b_f<b_{\max}<b_p$, the feasibility corridor is nondegenerate but the
coordination corridor is empty.

If another admissible distribution $\widetilde F$ satisfies
$\widetilde F(c)\geq F(c)$ for all $c$, both thresholds are weakly
lower and both corridors weakly larger. Multiplying every cost by
$\kappa>0$ multiplies $b_f,b_p$ by $\kappa$ and leaves $A_f,A_p$
unchanged.
\end{corollary}

The two ends of each corridor reflect different reasons a reference can
fail to attract users. With too little eligible debt, the market is too
thin for common use to be worthwhile. With too much, the claim's retained
payoff is diluted enough to outweigh the improvement in depth. A positive
network is possible only between these two extremes. Stationary selection
requires better quality than existence alone, so it moves both boundaries
inward.

Implementation costs matter at both stages. Lower costs make it easier
for a group of users to sustain its participation and improve the
potential of adoption relative to staying out. Both corridors therefore
expand when costs fall in the stated distributional sense, holding debt
quality fixed. Market infrastructure that lowers the recurring cost of using the claim widens both corridors, with debt and backing unchanged. A common rescaling of all costs changes how strong the
benchmark must be, but not the critical adoption shares: it changes the
level of the cost schedule without changing its relative shape.


Initial adoption creates a separate launch requirement. For
$A_0\in(0,A_f)$, the set of debt levels from which noiseless adoption
converges to the high equilibrium is
\[
 \{x:b(x;\varphi)>b_{\mathrm{initial\ share}}(A_0)\}.
\]
It is a nonempty open interval exactly when
$b_{\max}>b_{\mathrm{initial\ share}}(A_0)$, by
Corollary~\ref{cor:launch}. 

\subsection{Internalizing the adoption externality}
\label{sec:externality}

The analysis so far concerns private equilibria and their selection. We
now hold $x$, $\varphi$, and hence $b$ fixed and ask what an adoption
intervention can accomplish. This requires an explicit measure of the
benefits and costs of adoption, including allocations that need not be
private equilibria without the intervention.

\paragraph{Network surplus.}
At a cost-ordered allocation of size $A$, each adopter obtains gross
benefit $bA$, so aggregate net adoption surplus is
\begin{equation}
 U(A;b)=bA^2-\Ccal(A).
 \label{eq:surplus}
\end{equation}
For any fixed $A$, assigning adoption to the lowest-cost institutions
maximizes this surplus. Thus maximizing \eqref{eq:surplus} over $A$ is
also the optimum over all assignments of adoption to cost types. The
objective counts the institutions' network benefits and real
implementation costs; it holds debt and backing fixed and includes no
additional fiscal distortions or investor services.

At an interior cutoff allocation, the private gain of the marginal
adopter is $bA-F^{-1}(A)$, whereas the marginal change in aggregate surplus
is $U_A(A;b)=2bA-F^{-1}(A)$. The difference, $bA$, is the positive externality that the
additional adopter creates for institutions already using the reference.
This distinction also explains why a global potential comparison need
not coincide with a surplus comparison. The potential measures marginal
private incentives, not the sum of all adopters' payoffs.

Since $U(A;b)=\Pcal(A;2b)$, Theorem~\ref{thm:thresholds} immediately
gives the planner's allocation and the threshold for positive net gains:
\begin{equation}
 \begin{aligned}
 b_w&\equiv\min_{0<A<1}\frac{\Ccal(A)}{A^2}
       =\frac{b_p}{2}<b_f,\\
 \argmax_{0\leq A\leq1}U(A;b)&=\Mcal(2b).
 \end{aligned}
 \label{eq:welfarethreshold}
\end{equation}
A positive allocation with strictly positive net surplus exists exactly
when $b>b_w$. At $b=b_w$, both zero and $A_p$ attain surplus zero;
above it, the surplus-maximizing share is $A_+(2b)$.
The interval $(b_w,b_f)$ is therefore relevant to the intervention
problem: there are positive-surplus allocations there, but no positive
\emph{unsubsidized} adoption equilibrium. 

\paragraph{A matching transfer.}
Consider a policy paying each adopter
\begin{equation}
 \tau_\lambda(A)=\lambda bA,\qquad 0\leq\lambda\leq1.
 \label{eq:subsidy}
\end{equation}
Transfers are financed by nondistortionary lump-sum taxes on a passive
outside sector with sufficient resources. They are not drawn from the
backing resources $\varphi$. Including taxpayers, transfers cancel and
aggregate net surplus remains $U(A;b)$. The finance assumption separates
an adoption intervention from a change in benchmark quality.

The subsidized private gain is $(1+\lambda)bA-c$, and the induced
potential is
\begin{equation}
 \Pcal_\lambda(A;b)=\frac{(1+\lambda)b}{2}A^2-\Ccal(A).
 \label{eq:subsidizedpotential}
\end{equation}
In the finite game, adopter $i$ receives
$\lambda(b/n)\sum_{j\ne i}a_j$, with taxes paid by the outside sector.
The revision protocol is unchanged; its gains and potential replace $b$
by $(1+\lambda)b$. A policy therefore changes the incentives within the
same selection mechanism.

\begin{proposition}[Corrective transfers and implementation]
\label{prop:subsidy}
Under Assumption~\ref{ass:costs} and \eqref{eq:subsidy}:
\begin{enumerate}
 \item Positive private equilibria exist if and only if
 $(1+\lambda)b\geq b_f$. A positive potential maximizer exists if and only
 if $(1+\lambda)b\geq b_p$; it is the unique global maximizer when the
 latter inequality is strict.
 \item At $\lambda=1$, $\Pcal_1(A;b)=U(A;b)$ for every $A$.
 Stationary small-noise adoption converges to the unique surplus-maximizing
 share when $b\ne b_w$ and concentrates on $\{0,A_p\}$ when $b=b_w$.
\end{enumerate}
\end{proposition}

This matching transfer specializes the Pigouvian stochastic
implementation approach of \cite{Sandholm2007} to the adoption game.
The transfer rewards a benefit that an institution otherwise ignores.
Joining the reference helps not only the new adopter but also the users
who can now interact with it. At full correction, the payment equals
that additional benefit to existing users. Private adoption decisions
then reflect the marginal change in total network surplus. Because the
revision protocol selects according to the induced potential, aligning
that potential with surplus also aligns long-run selection with the
planner's adoption objective. When quality is high enough, even a rate below full correction can make
positive adoption a potential maximizer. The transfer changes the
relative attractiveness of using the reference and staying out. 

\paragraph{Launching versus maintaining a network.}
A matching transfer can help launch a network without being needed
permanently. Fix $b>b_f$ and consider noiseless renewal from an initial
share $A_0\in(0,A_-(b))$. Without intervention, this initial network
would shrink toward zero. Suppose a matching rate $\lambda\in(0,1]$
satisfies
\begin{equation}
 b>\frac{F^{-1}(A_0)}{(1+\lambda)A_0}.
 \label{eq:transferlaunch}
\end{equation}
By Corollary~\ref{cor:launch}, this condition places the initial share
in the basin of the high-adoption equilibrium at effective strength
$(1+\lambda)b$.

While the transfer remains in place, adoption increases toward
$A_+((1+\lambda)b)$. Since this subsidized equilibrium lies above
$A_f>A_-(b)$, the adoption share exceeds the original critical mass
$A_-(b)$ after a finite time. The government need not wait until the
subsidized equilibrium is reached. It can withdraw the transfer at
any finite date $T$ at which
\[
 A(T)>A_-(b).
\]
Thereafter, adoption follows the unsubsidized renewal law
$\dot A(t)=F(bA(t))-A(t)$. Proposition~\ref{prop:dynamics} therefore
implies
\[
 \lim_{t\to\infty}A(t)=A_+(b),
\]
without further transfers. 

This distinguishes launching a network from maintaining it. A small
initial network offers too few compatible counterparties to attract
additional users on its own. The transfer temporarily supplements
those benefits and helps build the installed base. Once participation
exceeds the unsubsidized critical mass, the network's own benefits
support convergence to the high-adoption equilibrium. Temporary
support can therefore change which equilibrium is reached without
changing the equilibrium that ultimately sustains the network.

\subsection{A conditional pricing implication}
\label{sec:pricing}

As a separate pricing extension, suppose passive risk-neutral investors
discount at $\beta\in(0,1)$. Each unit of the claim delivers the retained
real payoff $s(x;\varphi)$ and a next-period per-unit service $hA$, where
$h>0$. Its price is therefore
\begin{equation}
 p(x,A)=\beta\{s(x;\varphi)+hA\}.
 \label{eq:price}
\end{equation}
This extension takes adoption as determined by the unsubsidized game.
The additional investor service is not included in the adoption objective
of Section~\ref{sec:externality}; the equation values the network
conditional on adoption rather than changing institutions' revision
incentives.
Equation~\eqref{eq:price} is the cash-flow plus service-flow decomposition of \cite{BrunnermeierMerkelSannikov2024}, with the service proportional to the adoption share. The adoption game determines the quantity of service $A$; the parameter $h$ prices each unit and is taken as given. 
Write $b(x)=b(x;\varphi)$ at fixed $\varphi$.

Let $\Xcal(b_f)=[x_L(b_f),x_U(b_f)]$.
We consider an operating network continued along the high-adoption branch
up to $x_U(b_f)$. At that boundary its
adoption share approaches $A_f$. Immediately beyond the boundary, no
positive adoption equilibrium exists. 

\begin{corollary}[Loss of convenience value at the fold]
\label{cor:price}
Under Assumptions~\ref{ass:costs} and~\ref{ass:quality}, suppose
$b_{\max}>b_f$. The branch-continuation price loss at the
upper fold is
\begin{equation}
 \lim_{x\uparrow x_U(b_f)}p\bigl(x,A_+(b(x))\bigr)
 -\lim_{x\downarrow x_U(b_f)}p(x,0)
 =\beta hA_f.
 \label{eq:foldpriceloss}
\end{equation}
\end{corollary}

The loss comes from the disappearance of common use, rather than a
sudden reduction in the claim's retained cash payment. Along the
operating branch, investors value the services supplied by a positive
network. Once the claim no longer supports any positive equilibrium,
those services have no positive-equilibrium counterpart. Because payoff
retention is continuous at the boundary, the difference between the
limiting prices is exactly the discounted value of the lost service. \cite{HeNagelSong2022} explain the loss of Treasury convenience yields in March 2020 by dealer balance-sheet constraints. The fold here is a different mechanism, but it shares the feature that the service part of the price can disappear while the cash payoff does not change.

\section{Conclusion}
\label{sec:conclusion}

The limits of private coordination depend on whether the obstacle is
insufficient participation or insufficient benchmark quality. A claim
may support a functioning network without attracting enough initial
users to establish it. Conversely, a large installed base cannot
sustain an unsubsidized equilibrium once quality falls below the
existence threshold. The stationary logit comparison imposes a further
requirement on selection among feasible outcomes, but does not change
this underlying feasibility constraint.

This distinction gives temporary support a specific role. When a
self-sustaining equilibrium exists, a matching transfer that satisfies
the launch condition can help institutions build a sufficient user
base. Under noiseless renewal at fixed quality, support can then be
withdrawn once adoption exceeds the original critical mass. The
intervention changes which equilibrium is reached, rather than the
unsubsidized equilibrium itself. Successful launch therefore does not
remove the constraints that debt supply and backing place on continued
use.
The same launch problem applies to a proposed safe asset that does not yet have users, such as the senior tranche of pooled euro-area sovereign bonds in \cite{BrunnermeierEtAl2017}. Joint bonds issued by the European Union since 2020 face the same problem as a new issuer.

The analysis isolates these adoption constraints rather than solving
the issuer's financing problem. Their relevance extends beyond
participation: the price comparison at the existence boundary reflects
a loss of common-use services even though the retained cash payoff is
continuous.
A liquidity premium of the kind assumed in models of optimal public debt is available only inside the feasibility corridor, and it is the long-run outcome only inside the narrower coordination corridor. 


\newpage
\appendix
\section{Proofs}
\label{app:proofs}

Throughout the appendix, write $Q(u)=F^{-1}(u)$ for $0\leq u<1$,
with $Q(0)=0$, and let $R(c)=c/F(c)$ for $c>0$.
Both shorthand functions are used only in the proofs.

\subsection{Single-peaked costs, equilibria, and potential geometry}
\label{app:costs}

\begin{lemma}[Single-peaked density implies single crossing]
\label{lem:unimodality}
Under Assumption~\ref{ass:costs}, there is a unique $c_f>c_m$ such that
$c_f f(c_f)=F(c_f)$. Moreover,
\begin{equation}
 cf(c)-F(c)
 \begin{cases}
  >0,&0<c<c_f,\\
  =0,&c=c_f,\\
  <0,&c>c_f,
 \end{cases}
 \qquad f'(c_f)<0.
 \label{eq:singlecrossing}
\end{equation}
\end{lemma}

\begin{proof}
Let $D(c)=cf(c)-F(c)$, with $D(0)=0$. For $c>0$,
$D'(c)=cf'(c)$. Single-peakedness makes $D$ strictly increasing on
$(0,c_m)$ and strictly decreasing on $(c_m,\infty)$, with $D(c_m)>0$.
For sufficiently large $c$, the density is decreasing on $[c/2,c]$, so
\[
 \frac c2 f(c)\leq\int_{c/2}^c f(u)\dd u\longrightarrow0.
\]
Consequently $cf(c)\to0$ and $D(c)\to-1$. Continuity and strict
monotonicity on the decreasing part give a unique zero $c_f>c_m$ and
the signs in \eqref{eq:singlecrossing}. Since $c_f$ is above the mode,
$f'(c_f)<0$.
\end{proof}

\begin{proof}[Proof of Theorem~\ref{thm:equilibria} and its accompanying properties]
\leavevmode\par\nopagebreak
Since $f$ is continuous at zero and $f(0)=0$, $F(c)/c\to0$ as
$c\downarrow0$. Hence $R(c)\to\infty$ there. Also $R(c)\geq c$, so
$R(c)\to\infty$ as $c\to\infty$. Direct differentiation gives
\[
 R'(c)=\frac{F(c)-cf(c)}{F(c)^2}.
\]
Lemma~\ref{lem:unimodality} implies that $R$ strictly decreases up to
$c_f$ and strictly increases thereafter. Its unique minimizer is
therefore the cutoff defined in \eqref{eq:fold}. Changing variables
$A=F(c)$ gives the equivalent expression
\[
 b_f=\min_{0<A<1}\frac{Q(A)}{A},
\]
which we use in the potential and cost-distribution arguments below.

A positive equilibrium has $c=bA>0$, $A=F(c)$, and $R(c)=b$.
Conversely, every solution of $R(c)=b$ generates a positive equilibrium.
The shape and endpoint limits of $R$ give exactly one positive root at
$b_f$ and two for $b>b_f$, ordered as stated. They give the absence of
positive roots below $b_f$ as well. Zero solves the equilibrium equation
at every $b$.

For later arguments, note that the two positive cutoffs lie on opposite
sides of $c_f$, where $R'$ is nonzero. The implicit function theorem
therefore makes both cutoff branches, and hence both adoption branches,
continuously differentiable for $b>b_f$. Since $R$ diverges at both
endpoints and has the unique minimum $b_f$, both cutoffs converge to
$c_f$ as $b\downarrow b_f$. Thus both adoption shares converge to $A_f$.
These continuity and differentiability properties justify the envelope
comparison and the price limits used below; no formula for the branch
derivatives is needed.

For the potential claims, differentiation gives
\eqref{eq:potentialderivatives}. Since $Q(A)/A\to\infty$ at zero,
the derivative $\Pcal_A$ is negative in a right neighborhood of zero.
Thus zero is a strict one-sided local maximum. At $b>b_f$, the signs of
$\Pcal_A=bA-Q(A)$ on the intervals separated by $A_-$ and $A_+$
are respectively negative, positive, and negative. The lower root is a
strict local minimum, the upper root a strict local maximum, and these
are the only positive stationary points.

If $b\leq b_f$, $Q(A)\geq b_f A\geq bA$, with equality possible
only at $(A,b)=(A_f,b_f)$. Hence $\Pcal$ is strictly decreasing on
$[0,1)$, despite its zero derivative at $A_f$ when $b=b_f$. The fold
point is neither a local minimum nor a local maximum. Since
$\Pcal(0;b)=0$, zero is the unique global maximizer in this case.

Finally, for $A>F(b)$ one has $Q(A)>b\geq bA$, so the potential is
strictly decreasing beyond $F(b)$. Its global maximizers therefore lie
in $[0,F(b)]$, where it is finite and continuous. This proves existence
and compactness of $\Mcal(b)$, including when $\Ccal(1)=\infty$.
For $b>b_f$, the derivative signs already established show that only
zero and $A_+(b)$ can be global maximizers.
\end{proof}

\subsection{Noiseless basins and launch}

\begin{proof}[Proof of Proposition~\ref{prop:dynamics}]
Let $H(A;b)=F(bA)-A$. On $[0,1]$ it is continuously differentiable,
$H(0;b)=0$, and $H(1;b)=F(b)-1<0$, so the interval is forward
invariant and the differential equation has a unique solution.
At zero, $H_A(0;b)=bf(0)-1=-1$.
For $0<A<1$, strict monotonicity of $F$ gives
\[
 \operatorname{sgn}H(A;b)=\operatorname{sgn}(bA-Q(A)).
\]
The root count in Theorem~\ref{thm:equilibria} gives all basins by the
one-dimensional phase line. Below the fold, $H<0$ for all $A>0$.
Above it, the signs on the three intervals separated by $A_-$ and $A_+$
are negative, positive, and negative. Solutions are monotone toward the
stated limits, and uniqueness prevents crossing a stationary point.
At the fold, $H<0$ on both sides of $A_f$. Initial shares below $A_f$
therefore converge to zero, while those above it decrease toward $A_f$;
an initial share exactly equal to $A_f$ stays there.
\end{proof}

\begin{proof}[Proof of Corollary~\ref{cor:launch}]
Fix $A_0\in(0,A_f)$. On this interval,
$Q(A)/A=R(Q(A))$ decreases strictly from infinity to $b_f$.
The inequality $b>Q(A_0)/A_0$ is equivalent to $F(bA_0)>A_0$.
It holds exactly when $A_-(b)<A_0<A_+(b)$; here $A_0<A_f$ ensures
that the upper inequality is automatic once the positive roots exist.
Proposition~\ref{prop:dynamics} gives convergence to the high root.
At equality, $A_0$ is the lower stationary point. For
$b_f<b<b_{\mathrm{initial\ share}}(A_0)$, the initial share lies below
$A_-(b)$ and converges to zero. For $b\leq b_f$, convergence to zero
follows from $A_0<A_f$ and Proposition~\ref{prop:dynamics}. This proves
necessity and sufficiency. The shape and endpoint limit of $Q(A)/A$ establish the
comparative statics of the threshold in \eqref{eq:initialshare}.
\end{proof}

\subsection{The invariant distribution and its concentration}

\begin{proof}[Proof of Theorem~\ref{thm:logit} and the expected-potential
limit]
\emph{Step 1: exact potential and invariant probabilities.}
If $a_i$ changes from zero to one, with others fixed, the change in
$\Phi_n$ is
$(b/n)\sum_{j\ne i}a_j-c_{i,n}=g_{i,n}(\mathbf{a}_{-i})$.
For adjacent profiles $\mathbf{a},\mathbf{a}'$ differing by that change,
let $q_{n,\eta}(\mathbf{a},\mathbf{a}')$ denote the continuous-time
transition rate. Unit-rate individual revision gives
\[
 \frac{q_{n,\eta}(\mathbf{a},\mathbf{a}')}
      {q_{n,\eta}(\mathbf{a}',\mathbf{a})}
 =\frac{L(g_{i,n}(\mathbf{a}_{-i});\eta)}
        {1-L(g_{i,n}(\mathbf{a}_{-i});\eta)}
 =\exp\!\left\{\frac{\Phi_n(\mathbf{a}';b)
                         -\Phi_n(\mathbf{a};b)}{\eta}\right\}.
\]
Thus \eqref{eq:gibbs} satisfies detailed balance. Positive rates for
all single-coordinate changes establish irreducibility, so the finite
continuous-time chain has this unique invariant distribution and
converges to it from every initial profile.

The chain observed at revision opportunities chooses one institution
uniformly, including opportunities that leave its action unchanged.
If its transition kernel is $K_{n,\eta}$, the continuous-time generator
is $n(K_{n,\eta}-I)$. Hence the two processes have identical invariant
probabilities, as stated in Section~\ref{sec:finite}.

\emph{Step 2: approximation of the best potential at each share.}
At a fixed number $k$ of adopters, network benefits depend only on $k$,
so the potential is highest when the $k$ lowest costs adopt. Its
normalized value is
\begin{equation}
 \widehat{\Pcal}_n(k)=\frac b2\frac{k}{n}\left(\frac{k}{n}-\frac1n\right)-C_n(k),
 \qquad C_n(k)=\frac1n\sum_{i=1}^k Q\!\left(\frac{i}{n+1}\right).
 \label{eq:discretepotential}
\end{equation}
Let $Q_n(u)=Q(\lceil nu\rceil/(n+1))$ for $u\in(0,1]$ and
$Q_n(0)=0$. Then $C_n(k)=\int_0^{k/n}Q_n(u)\dd u$, and
$Q_n\to Q$ uniformly on compact subintervals of $[0,1)$.
For any sequence $k_n/n\to A<1$, uniform convergence and boundedness
on a slightly larger compact interval imply
$C_n(k_n)\to\Ccal(A)$ and $\widehat{\Pcal}_n(k_n)\to\Pcal(A;b)$.
For a sequence with limit $A=1$, nonnegativity and Fatou's lemma give
\[
 \liminf_n C_n(k_n)
 =\liminf_n\int_0^1\mathbf{1}\{u\leq k_n/n\}Q_n(u)\dd u
 \geq\Ccal(1).
\]
Consequently, for every limiting share $A\in[0,1]$,
\begin{equation}
 \limsup_n \widehat{\Pcal}_n(k_n)\leq\Pcal(A;b)
 \quad\text{whenever }k_n/n\to A,
 \label{eq:upperlimit}
\end{equation}
with equality for $A<1$. This covers infinite mean costs as well.

\emph{Step 3: a potential gap and concentration.}
The extended-valued potential is upper semicontinuous on $[0,1]$ and
has finite maximum $G=\max_A\Pcal(A;b)$. As established after
Theorem~\ref{thm:equilibria}, its maximizing set is nonempty, compact,
and contained in $[0,F(b)]$. Approximating a maximizer by $k_n/n$ gives
$\liminf_n\max_k \widehat{\Pcal}_n(k)\geq G$. Compactness and
\eqref{eq:upperlimit} give the reverse inequality, so
$\max_k \widehat{\Pcal}_n(k)\to G$.

Fix $\delta>0$ and let
$D_\delta=\{A\in[0,1]:\dist(A,\Mcal(b))\geq\delta\}$.
If this set is empty, concentration is immediate. Otherwise,
upper semicontinuity on the compact set gives a strict gap: for some
$g>0$, $\Pcal(A;b)\leq G-g$ on $D_\delta$.
If all the values there are $-\infty$, any finite $g>0$ suffices.
A subsequence argument using \eqref{eq:upperlimit} shows that, for all
sufficiently large $n$,
\[
 \max_{k:k/n\in D_\delta}\widehat{\Pcal}_n(k)\leq G-g/2,
 \qquad \max_k \widehat{\Pcal}_n(k)\geq G-g/4,
\]
where an empty constrained maximum can be omitted. Every profile has
potential no greater than the lowest-cost profile with the same share.
Thus the unnormalized potential gap between any profile with adoption
in $D_\delta$ and a global maximizing profile is at least $ng/4$.
There are at most $2^n$ profiles. Bounding the Gibbs numerator by that
count and the denominator by a single maximizing profile yields
\[
 \mu_{n,\eta_n}\!\left(A_n(\mathbf{a})\in D_\delta\right)
 \leq\exp\!\left\{n\log2-\frac{ng}{4\eta_n}\right\}
 \longrightarrow0.
\]
This proves \eqref{eq:concentration} without using the later
characterization of $\Mcal(b)$.

\emph{Step 4: expected potential.}
Let $Z_{n,\eta}$ be the denominator in \eqref{eq:gibbs}, and let
\[
 S_{n,\eta}=-\sum_{\mathbf{a}\in\{0,1\}^n}
 \mu_{n,\eta}(\mathbf{a})\log\mu_{n,\eta}(\mathbf{a})
\]
be the entropy of the invariant distribution. The Gibbs formula gives
$\log Z_{n,\eta}=\E_{\mu}[\Phi_n]/\eta+S_{n,\eta}$.
Since $\log Z_{n,\eta}\geq\max_{\mathbf{a}}\Phi_n(\mathbf{a};b)/\eta$
and $S_{n,\eta}\leq n\log2$,
\[
 0\leq\max_k \widehat{\Pcal}_n(k)
      -\E_{\mu_{n,\eta}}[\Phi_n(\mathbf{a};b)/n]
 \leq\eta\log2.
\]
Together with $\max_k \widehat{\Pcal}_n(k)\to G$, this proves the expected-potential
limit stated after the theorem.
\end{proof}

\subsection{Finite-horizon consistency of the population descriptions}
\label{app:finitehorizon}

The renewal laws in Section~\ref{sec:renewal} are defined directly from
continuum revision flows. The following result verifies that they also
approximate the finite system in Section~\ref{sec:finite} on every fixed
time horizon. It supplies the connection used in the order-of-limits
discussion, rather than a prerequisite for the continuum formulation or
the stationary selection theorem. Initial profiles may have any cost
composition; only their aggregate shares are required to converge.

For this subsection, let $\mathbf{a}_{n,\eta}(t)$ denote the
action-profile process generated by the revision protocol in
Section~\ref{sec:finite}, and write
\[
 A_{n,\eta}(t)=A_n\bigl(\mathbf{a}_{n,\eta}(t)\bigr)
\]
for its aggregate adoption share at time $t$.
\begin{lemma}[Finite-horizon renewal limit]
\label{lem:finitehorizon}
Maintain Assumption~\ref{ass:costs} and the finite-population protocol in
Section~\ref{sec:finite}. Fix $b>0$, $A_0\in[0,1]$, and a finite horizon
$T>0$. For fixed $\eta>0$, suppose $A_{n,\eta}(0)\xrightarrow{p}A_0$.
Then
\[
 \sup_{0\leq t\leq T}|A_{n,\eta}(t)-A(t;\eta)|
 \xrightarrow{p}0,
\]
where $A(\cdot;\eta)$ solves \eqref{eq:noisyrenewal} with
$A(0;\eta)=A_0$.
For every sequence $\eta_n>0$ with $\eta_n\to0$, if
$A_{n,\eta_n}(0)\xrightarrow{p}A_0$, then
\[
 \sup_{0\leq t\leq T}|A_{n,\eta_n}(t)-A(t)|
 \xrightarrow{p}0,
\]
where $A$ solves \eqref{eq:renewal} with initial share $A_0$.
Also, $A(\cdot;\eta)\to A(\cdot)$ uniformly on $[0,T]$ as
$\eta\downarrow0$ when both paths start at $A_0$.
\end{lemma}

\begin{proof}
Let $\bar f=f(c_m)<\infty$, the upper bound on the density, and extend
$F$ to be zero on negative arguments. Let $F_n$ be the empirical CDF of
the quantile costs in \eqref{eq:finitecosts}. Their construction implies
\[
 \varepsilon_n\equiv\sup_{c\in\R}|F_n(c)-F(c)|\leq\frac1n.
\]
For the proof, define $S_\eta(A)=\int L(bA-c;\eta)\dd F(c)$ for
real $A$ and $H_\eta(A;b)=S_\eta(A)-A$. If $Z$ has CDF
$L(y;1)=(1+e^{-y})^{-1}$, then
\[
 S_\eta(A)=\E[F(bA-\eta Z)].
\]
The extended CDF is globally $\bar f$-Lipschitz. Thus $S_\eta$ is
$b\bar f$-Lipschitz, uniformly in $\eta$, and
\begin{equation}
 \sup_{A\in[0,1]}|S_\eta(A)-F(bA)|
 \leq \bar f\eta\E|Z|=2\bar f\eta\log2.
 \label{eq:smoothingbound}
\end{equation}
Here $\E|Z|=2\int_0^\infty(1+e^y)^{-1}\dd y=2\log2$.
Consequently both $H_\eta$ and $H_0(A;b)=F(bA)-A$ are
$(1+b\bar f)$-Lipschitz. Their vector fields point into $[0,1]$ at
its endpoints, so their solutions exist uniquely and remain in that
interval.

At profile $\mathbf{a}$, the exact conditional drift of aggregate
adoption is
\[
 d_{n,\eta}(\mathbf{a})
 =\frac1n\sum_{i=1}^n
 \left[L\!\left(b\left(A_n(\mathbf{a})-\frac{a_i}{n}\right)
                   -c_{i,n};\eta\right)-a_i\right].
\]
For any real $A$, integration by parts against the bounded monotone
function $c\mapsto L(bA-c;\eta)$ gives
\[
 \left|\int L(bA-c;\eta)\dd F_n(c)-S_\eta(A)\right|
 \leq\varepsilon_n.
\]
For $A=A_n(\mathbf{a})$, the average of the logit terms in the drift
is bounded above by the empirical expression at $A$ and below by that
expression at $A-1/n$. The uniform Lipschitz bound for $S_\eta$ then
implies
\begin{equation}
 \sup_{\mathbf{a}}\left|
 d_{n,\eta}(\mathbf{a})-H_\eta(A_n(\mathbf{a});b)\right|
 \leq\varepsilon_n+\frac{b\bar f}{n}.
 \label{eq:driftbound}
\end{equation}
Combining \eqref{eq:smoothingbound} and \eqref{eq:driftbound} gives a
bound relative to the noiseless drift that is uniform in the profile:
\[
 \sup_{\mathbf{a}}\left|
 d_{n,\eta}(\mathbf{a})-H_0(A_n(\mathbf{a});b)\right|
 \leq\varepsilon_n+\frac{b\bar f}{n}+2\bar f\eta\log2.
\]

The aggregate finite process has the decomposition
\[
 A_{n,\eta}(t)=A_{n,\eta}(0)
 +\int_0^t d_{n,\eta}(\mathbf{a}_{n,\eta}(s))\dd s+M_{n,\eta}(t),
\]
where $M_{n,\eta}$ is a martingale starting at zero. An aggregate jump
has magnitude at most $1/n$, and total revision opportunities arrive at
rate $n$. Its expected quadratic variation up to $T$ is therefore at
most $T/n$, and the martingale maximal inequality gives
\[
 \E\!\left[\sup_{0\leq t\leq T}|M_{n,\eta}(t)|^2\right]
 \leq\frac{4T}{n}.
\]
Comparing the decomposition with the integral equation for $A(\cdot;\eta)$
and applying Gronwall's inequality yields
\[
 \begin{split}
 \sup_{t\leq T}|A_{n,\eta}(t)-A(t;\eta)|
 \leq e^{(1+b\bar f)T}\biggl(&|A_{n,\eta}(0)-A_0|
       +T\left(\varepsilon_n+\frac{b\bar f}{n}\right)\\
       &+\sup_{t\leq T}|M_{n,\eta}(t)|\biggr).
 \end{split}
\]
This proves the fixed-noise approximation. Using the noiseless drift
comparison instead adds $2T\bar f\eta_n\log2$ to the bound and proves
the joint large-population, vanishing-noise result for any such sequence.
Finally, the same deterministic comparison gives
\[
 \sup_{t\leq T}|A(t;\eta)-A(t)|
 \leq 2T\bar f\eta\log2\,e^{(1+b\bar f)T},
\]
proving the last claim. All statements concern fixed $T$; none exchanges
this limit with an unbounded time horizon.
\end{proof}

\subsection{The equilibrium-selection threshold}

\begin{proof}[Proof of Theorem~\ref{thm:thresholds} and
\eqref{eq:thresholdbounds}]
Section~\ref{sec:equilibria} established that zero is the unique
potential maximizer when $b\leq b_f$. For $b>b_f$, the only candidates
are zero and $A_+(b)$. Let
$V(b)=\Pcal(A_+(b);b)$ and extend it continuously to $b_f$ by
$V(b_f)=\Pcal(A_f;b_f)$. Since $Q(u)>b_f u$ on $(0,A_f)$,
\[
 V(b_f)=\frac{b_f A_f^2}{2}-\int_0^{A_f}Q(u)\dd u<0.
\]
For $b>b_f$, the first-order condition at $A_+(b)$ gives the envelope
derivative
\[
 V'(b)=\frac{A_+(b)^2}{2}\geq\frac{A_f^2}{2}>0.
\]
Thus $V(b)\to\infty$ and has a unique zero $b_p>b_f$.
Its sign gives \eqref{eq:selection}. The positive maximizer at equality
is $A_p=A_+(b_p)>A_f$.

At $b_p$, the potential is nonpositive for every share, with equality
at zero and the unique positive maximizer $A_p$. For every $A>0$,
\[
 \frac{2\Ccal(A)}{A^2}\geq b_p,
\]
with equality exactly at $A_p$. This proves \eqref{eq:bp} and uniqueness
of its minimizer. Finally, strict monotonicity of $Q$ implies
\[
 \Ccal(A_f)<A_fQ(A_f)=b_f A_f^2.
\]
Evaluating the minimum in \eqref{eq:bp} at $A_f$ therefore gives
$b_p\leq2\Ccal(A_f)/A_f^2<2b_f$. This proves
\eqref{eq:thresholdbounds}. The stationary-adoption statements follow
by applying Theorem~\ref{thm:logit} to the characterized maximizing set.
\end{proof}

\subsection{Debt corridors and comparative statics}

\begin{proof}[Proof of Corollary~\ref{col:corridors}]
\leavevmode\par\nopagebreak
Under Assumption~\ref{ass:quality}, since $b_p>b_f$, $\Xcal_p\subseteq\Xcal_f$.
If $b_{\max}>b_p$, raising the threshold from $b_f$ to $b_p$ moves
both endpoints strictly inward, so the inclusion is strict. The case
$b_f<b_{\max}<b_p$ follows directly from the level-set description.

Corollary~\ref{cor:launch} gives the launch requirement;
strict single-peakedness gives its open-interval form whenever peak
strength strictly exceeds the initial-share threshold.

If $\widetilde F\geq F$, then $\widetilde Q\leq Q$ and
$\widetilde{\Ccal}\leq\Ccal$. Taking minima in \eqref{eq:fold} and
\eqref{eq:bp} gives weakly lower existence and selection thresholds,
whose superlevel sets therefore expand. Scaling costs by $\kappa$
scales $Q$ and $\Ccal$ by $\kappa$, without changing the minimizing
shares in those two formulas. Hence $b_f,b_p$ scale by $\kappa$, while
$A_f,A_p$ remain unchanged.
\end{proof}

\subsection{Network surplus and corrective transfers}

\begin{proof}[The surplus identities in Section~\ref{sec:externality}]
For a fixed share $A$, total gross network benefit is $bA^2$ regardless
of which institutions adopt. Lowest-cost assignment minimizes its
implementation cost, giving \eqref{eq:surplus}. The identity
$U(A;b)=\Pcal(A;2b)$ and Theorem~\ref{thm:thresholds} imply
\eqref{eq:welfarethreshold} and the complete maximizing-set description.
The positive-surplus condition also follows directly from
$U(A;b)=A^2[b-\Ccal(A)/A^2]$ for $A>0$.
At $b=b_w$, zero and $A_p$ attain the maximum value zero; above this
threshold, $A_+(2b)$ uniquely maximizes surplus at a strictly positive
value. Since $b_p<2b_f$, $b_w<b_f$. Finally, differentiating
\eqref{eq:surplus} and subtracting \eqref{eq:potentialderivatives}
gives the externality $bA$ at the marginal adopter.
\end{proof}

\begin{proof}[Proof of Proposition~\ref{prop:subsidy} and the launch comparison]
Outside taxes do not depend on an institution's action. Adding
\eqref{eq:subsidy} therefore replaces $b$ by $(1+\lambda)b$ in the
private equilibrium equation and in the finite-player gain.
Theorems~\ref{thm:equilibria}, \ref{thm:logit}, and~\ref{thm:thresholds}
give the existence and selection conditions with the original costs.

Transfers and outside taxes cancel in aggregate. At $\lambda=1$,
\eqref{eq:subsidizedpotential} equals $U(A;b)$ for every share. The
same identity holds at the finite-profile level: the full-correction
potential is
\[
 \Phi_n(\mathbf{a};2b)
 =\frac{2b}{n}\sum_{i<j}a_i a_j-\sum_i c_{i,n}a_i
 =\sum_i a_i\left(\frac bn\sum_{j\ne i}a_j-c_{i,n}\right),
\]
which is the sum of the unsubsidized institutional payoffs, and hence
net surplus after cancelling the policy transfers. Applying the
concentration and expected-potential conclusions of
Theorem~\ref{thm:logit} at strength $2b$ shows that stationary aggregate
adoption concentrates on $\argmax_A U(A;b)$ and stationary expected
surplus per institution tends to its maximum. In particular, this
conclusion accounts for the cost composition of adoption, not only its
aggregate share. The maximizing sets at, above, and below $b_w$ follow
from \eqref{eq:welfarethreshold}; no mixture weights at equality are
asserted.

For the launch comparison, holding the transfer in place replaces $b$ by
$(1+\lambda)b$ in the noiseless renewal law as well. For $A_0\in(0,A_f)$,
Corollary~\ref{cor:launch} therefore gives convergence to the subsidized
high equilibrium exactly when
\[
 (1+\lambda)b>\frac{Q(A_0)}{A_0},
\]
which is \eqref{eq:transferlaunch}. At equality, $A_0$ is the subsidized
lower equilibrium rather than a point in the high equilibrium's basin.
For a fixed $b>b_f$, the subsidized critical initial share is
$A_-((1+\lambda)b)$. The function $Q(A)/A=R(Q(A))$ is strictly decreasing
on $(0,A_f)$, so its lower-branch solution falls as $\lambda$ rises. This
proves the basin expansion. At $A_0=0$, the renewal drift remains zero
for every $\lambda$ because $F(0)=0$.
\end{proof}

\subsection{Convenience value}

\begin{proof}[Proof of Corollary~\ref{cor:price}]
By Assumption~\ref{ass:quality} and $b_{\max}>b_f$, the feasibility
corridor is a nondegenerate compact interval. Its upper endpoint
satisfies $b(x_U(b_f);\varphi)=b_f$. Approaching this boundary from within the
feasibility corridor gives $b(x)\downarrow b_f$ and
$A_+(b(x))\to A_f$, by the continuity established in the proof of
Theorem~\ref{thm:equilibria}. Immediately to its right, $b(x)<b_f$,
so zero is the only adoption equilibrium. Payoff retention
$s(x;\varphi)$ is continuous at $x_U(b_f)$. Subtracting the two
one-sided limits of \eqref{eq:price} therefore cancels the retained
cash payoff and leaves $\beta hA_f$, as in \eqref{eq:foldpriceloss}.
\end{proof}

\section{Allowing positive density at zero}
\label{app:positive_density_zero}

This section relaxes the requirement that the cost density vanish at
zero. We retain all other conditions in Assumption~\ref{ass:costs},
including the strictly interior mode $c_m>0$ and the strict increase
and decrease of the density on either side of that mode. We focus on
the additional case $0<f(0)<\infty$, since $f(0)=0$ is covered by the
main analysis. The condition $F(0)=0$ is maintained: a positive density
at zero does not introduce a positive mass of costless institutions.

Define
\begin{equation}
 b_0\equiv\frac{1}{f(0)}.
 \label{eq:positive_density_b0}
\end{equation}
This additional threshold determines whether a sufficiently small
positive network expands under noiseless renewal. The existence
threshold and the stationary-selection threshold remain distinct,
and both lie strictly below $b_0$.

\subsection{Equilibrium existence and stationary selection}\label{sec: density extension}

The definitions of $c_f$, $b_f$, and $A_f$ in \eqref{eq:fold}
remain well defined. The only change in the positive-equilibrium
count is that the lower branch now reaches zero at a finite
benchmark strength.

\begin{proposition}[Positive density at zero]
\label{prop:positive_density_zero}
Under the conditions of this section:
\begin{enumerate}
 \item The function $c/F(c)$ has a unique interior minimizer $c_f$,
 with $c_f>c_m$ and $b_f<b_0$. Zero adoption is always an equilibrium.
 The positive equilibria are
 \[
 \{A\in(0,1):A=F(bA)\}
 =
 \begin{cases}
  \varnothing, & 0<b<b_f,\\
  \{A_f\}, & b=b_f,\\
  \{A_-(b),A_+(b)\}, & b_f<b<b_0,\\
  \{A_+(b)\}, & b\geq b_0.
 \end{cases}
 \]
 The lower branch satisfies $A_-(b)\to0$ as $b\uparrow b_0$.
 The upper branch is defined for every $b>b_f$ and satisfies
 $A_+(b)\to A_f$ as $b\downarrow b_f$.

 \item The potential-maximizer characterization in
 \eqref{eq:selection} remains valid. Its threshold is
 \[
 b_p=\min_{0<A<1}\frac{2\Ccal(A)}{A^2},
 \]
 with the minimum attained uniquely at $A_p=A_+(b_p)$. Moreover,
 \begin{equation}
  b_f<b_p<\min\{2b_f,b_0\},
  \qquad A_p>A_f.
  \label{eq:positive_density_ordering}
 \end{equation}
\end{enumerate}
\end{proposition}

\begin{proof}
Write $R(c)=c/F(c)$ and $D(c)=cf(c)-F(c)$. Continuity of $f$
at zero gives
\[
 \lim_{c\downarrow0}\frac{F(c)}{c}=f(0),
 \qquad
 \lim_{c\downarrow0}R(c)=b_0.
\]
Also, $R(c)\to\infty$ as $c\to\infty$.

The single-crossing argument does not require $f(0)=0$.
Indeed, $D(0)=0$ and
\[
 D'(c)=cf'(c).
\]
Thus $D$ increases strictly up to $c_m$ and decreases strictly
thereafter. Since the density is eventually decreasing and
integrable, $cf(c)\to0$, so $D(c)\to-1$. There is consequently a
unique $c_f>c_m$ with $D(c_f)=0$. Because
\[
 R'(c)=-\frac{D(c)}{F(c)^2},
\]
$R$ decreases strictly from its limiting value $b_0$ to $b_f$
on $(0,c_f)$ and increases strictly from $b_f$ to infinity on
$(c_f,\infty)$. In particular, $b_f<b_0$.

Positive adoption equilibria correspond exactly to solutions of
$R(c)=b$, with adoption share $F(c)$. The decreasing part of $R$
takes precisely the values in $(b_f,b_0)$, whereas its increasing
part takes all values above $b_f$. This proves the root count and
the branch limits in part (i). The fold remains nondegenerate
because $f'(c_f)<0$.

For part (ii), write $Q=F^{-1}$. The potential derivative remains
\[
 \Pcal_A(A;b)=bA-Q(A).
\]
For $b\leq b_f$, it is negative except possibly at $A_f$ when
$b=b_f$, so zero is the unique global maximizer. For
$b_f<b<b_0$, the potential decreases, increases, and then decreases
on the intervals separated by $A_-(b)$ and $A_+(b)$. Its only
global candidates are therefore zero and $A_+(b)$. For $b\geq b_0$,
the potential increases on $(0,A_+(b))$ and decreases thereafter,
so $A_+(b)$ is its unique global maximizer.

Define
\[
 V(b)=\Pcal(A_+(b);b),\qquad b>b_f,
\]
and extend it continuously to $b_f$. Since $Q(u)>b_f u$ on
$(0,A_f)$,
\[
 V(b_f)
 =\frac{b_fA_f^2}{2}-\int_0^{A_f}Q(u)\dd u<0.
\]
Along the upper branch, the first-order condition gives
\[
 V'(b)=\frac{A_+(b)^2}{2}>0.
\]

To locate the zero of $V$ relative to $b_0$, use the strict increase
of $f$ near the origin. For $0<c<c_m$,
\[
 F(c)=\int_0^c f(u)\dd u>f(0)c.
\]
Equivalently, $Q(A)<b_0A$ for $0<A<F(c_m)$. Hence, for every such
$A$,
\[
 \Pcal(A;b_0)
 =\frac{b_0A^2}{2}-\int_0^A Q(u)\dd u>0.
\]
It follows that $V(b_0)>0$. Continuity and strict monotonicity
therefore give a unique $b_p\in(b_f,b_0)$ at which $V(b_p)=0$.
The sign of $V$ establishes \eqref{eq:selection}, and
$A_p=A_+(b_p)>A_f$.

At $b_p$, the potential is nonpositive everywhere and equals zero
only at $0$ and $A_p$. Thus
\[
 \frac{2\Ccal(A)}{A^2}\geq b_p
 \quad\text{for every }A\in(0,1),
\]
with equality only at $A_p$. This proves the stated characterization
of $b_p$. Finally, since $Q$ is strictly increasing,
\[
 \Ccal(A_f)<A_fQ(A_f)=b_fA_f^2.
\]
Evaluating the minimum defining $b_p$ at $A_f$ gives
$b_p<2b_f$, completing \eqref{eq:positive_density_ordering}.
\end{proof}

The substantive existence--selection distinction is therefore
unchanged. A positive network first becomes feasible at $b_f$,
but it wins the global potential comparison only at the higher
threshold $b_p$. In particular, the whole interval $(b_f,b_p)$
lies below $b_0$, where zero and high adoption are both locally
persistent. Positive density at zero changes what happens at
sufficiently high quality, not the coexistence underlying this gap.

\subsection{Noiseless adjustment and launching}

The renewal law remains
\[
 \dot A=F(bA)-A.
\]
Its derivative at zero is now
\[
 \left.\frac{\partial}{\partial A}
       \bigl[F(bA)-A\bigr]\right|_{A=0}
 =bf(0)-1.
\]
Zero adoption is therefore locally asymptotically stable when
$b<b_0$ and unstable when $b>b_0$. At $b=b_0$, the derivative is
zero, but strict increase of the density near the origin gives
\[
 F(b_0A)>f(0)b_0A=A
\]
for all sufficiently small $A>0$. Thus zero is unstable from the
right at $b=b_0$ as well. Correspondingly, zero is a strict
one-sided local potential maximum for $b<b_0$ and a strict
one-sided local minimum for $b\geq b_0$.

For $b<b_f$ and at $b=b_f$, the convergence conclusions in
Proposition~\ref{prop:dynamics} are unchanged. For
$b_f<b<b_0$, its two-basin description also remains unchanged:
initial adoption below $A_-(b)$ converges to zero, while initial
adoption above $A_-(b)$ converges to $A_+(b)$.
For $b\geq b_0$, however, the lower positive equilibrium is absent
and
\[
 A(0)>0
 \quad\Longrightarrow\quad
 \lim_{t\to\infty}A(t)=A_+(b).
\]
Indeed, the renewal drift is positive below the unique positive
root and negative above it. An initial share exactly equal to
zero still stays at zero because $F(0)=0$. Instability means that
a small positive initial share grows, not that the noiseless
process leaves zero spontaneously.

The launch condition in Corollary~\ref{cor:launch} retains its
form. For $A_0\in(0,A_f)$, convergence to the high equilibrium
occurs if and only if
\[
 b>\frac{F^{-1}(A_0)}{A_0}.
\]
This threshold decreases strictly with $A_0$, but its limit is
now finite:
\begin{equation}
 b_f<
 \frac{F^{-1}(A_0)}{A_0}
 <b_0,
 \qquad
 \lim_{A_0\downarrow0}
 \frac{F^{-1}(A_0)}{A_0}=b_0.
 \label{eq:positive_density_launch}
\end{equation}
These conclusions follow from the decreasing part of $R$ and
the phase-line argument above. At equality, the initial share is
the lower equilibrium and does not converge to the high one.

Economically, a positive density at zero supplies enough very
low-cost institutions for an arbitrarily small positive network
to expand when quality is sufficiently high. The assumption
$f(0)=0$ excludes this possibility at every finite quality.
It is useful for obtaining a positive critical mass throughout
the high-quality region, but is not necessary for a critical
mass near the existence and selection thresholds.

\subsection{Implications for the remaining results}

\paragraph{Stochastic selection and population approximation.}
Theorem~\ref{thm:logit} continues to hold under the relaxed
assumptions. The exact-potential identity, detailed balance, and
Gibbs invariant distribution do not use $f(0)=0$. The
large-population concentration proof uses the continuity of the
inverse distribution, the approximation of cutoff costs, and a
potential gap away from the global maximizing set. These
properties remain valid. In particular, the potential still
attains its maximum in $[0,F(b)]$, and the proof accommodates
infinite mean costs.

Lemma~\ref{lem:finitehorizon} is unchanged as well. Its drift and
smoothing estimates use the bounded density
$\sup_c f(c)=f(c_m)$, not a zero density at the origin.
Thus the continuum renewal law and the finite-population
stationary comparison retain their common revision foundation.

\paragraph{Debt corridors, surplus, and corrective transfers.}
The feasibility and coordination corridors remain determined by
$b_f$ and $b_p$, respectively. Their nesting and the
cost-distribution comparative statics follow from the same
threshold formulas. Under the maintained quality and pricing
conditions, the branch-continuation price loss at the upper fold
also remains $\beta hA_f$: the positive branch still ends at the
strictly positive share $A_f$.

The identity $U(A;b)=\Pcal(A;2b)$ is unaffected, so the social
threshold remains
\[
 b_w=\frac{b_p}{2}<b_f.
\]
A matching transfer still replaces private strength $b$ by
$(1+\lambda)b$, and full correction still aligns the potential
with network surplus. Consequently, the existence and
stationary-implementation conclusions of
Proposition~\ref{prop:subsidy} remain valid.

Temporary support can still help launch a self-sustaining
network in the coexistence region $b_f<b<b_0$. Starting from
$A_0\in(0,A_-(b))$, any admissible matching rate satisfying
\[
 (1+\lambda)b>\frac{F^{-1}(A_0)}{A_0}
\]
makes adoption grow toward the subsidized high equilibrium.
That equilibrium lies above $A_f>A_-(b)$, so adoption crosses
the unsubsidized critical mass after a finite time. Withdrawing
support strictly after that crossing leads to the unsubsidized
high equilibrium, as in the main analysis. For $b\geq b_0$,
by contrast, every strictly positive initial share already
converges to the high equilibrium without a transfer.

\bibliographystyle{apalike}
\bibliography{references}
\end{document}